\documentclass[12pt,letterpaper,top=2.5cm,bottom=2.5cm,left=3cm,right=2.5cm,marginparwidth=1.75cm]{article}
\usepackage{arxiv}

\usepackage{graphicx}

\usepackage{caption}
\usepackage{subcaption}
\usepackage{textcomp}
\usepackage{url}
\usepackage{colortbl}
\usepackage{soul}
\usepackage{multirow}
\usepackage{float}
\usepackage{physics}
\usepackage{pifont}
\usepackage{color}
\usepackage{alltt}
\usepackage[hidelinks]{hyperref}
\usepackage{enumerate}
\usepackage{siunitx}
\usepackage{breakurl}
\usepackage{epstopdf}
\usepackage{pbox}
\usepackage{stfloats}
\usepackage[vlined,linesnumberedhidden,ruled,resetcount]{algorithm2e}
\usepackage{algpseudocode}  
\usepackage{cite}
\usepackage{amsmath, amsfonts}\usepackage{graphicx}

\usepackage{caption}
\usepackage{subcaption}
\usepackage{textcomp}
\usepackage{url}
\usepackage{colortbl}
\usepackage{multirow}
\usepackage{float}
\usepackage{physics}
\usepackage{pifont}
\usepackage{color}
\usepackage{alltt}
\usepackage[hidelinks]{hyperref}
\usepackage{enumerate}
\usepackage{siunitx}
\usepackage{breakurl}
\usepackage{epstopdf}
\usepackage{pbox}
\usepackage{stfloats}
\usepackage[vlined,linesnumberedhidden,ruled,resetcount]{algorithm2e}
\usepackage{algpseudocode}  
\usepackage{cite}
\usepackage{amsmath, amsfonts}

\allowdisplaybreaks

\usepackage{titlesec}
\usepackage{array}     % m{<width>} column type
\usepackage{diagbox}   % diagonal header cell
\usepackage{booktabs}  % optional, for nicer rules

\begin{document}
%\begin{frontmatter}

\title{Physics-Informed Dynamical Modeling of Extrusion-Based 3D Printing Processes} 

\author{Mandana Mohammadi Looey, 
Marissa Loraine Scalise, 
Amrita Basak, 
and Satadru Dey
\thanks{The authors are with the Department of Mechanical Engineering, The Pennsylvania State University, University Park, Pennsylvania 16802, USA. (e-mails: mfm6970@psu.edu, mls7478@psu.edu, aub1526@psu.edu, skd5685@psu.edu).}
\thanks{This work was supported by National Science Foundation under Grants No. 2346650. The opinions, findings, and conclusions or recommendations expressed are those of the author(s) and do not necessarily reflect the views of the National Science Foundation.}
}

\maketitle
% \address[Second]{Colorado State University, 
%    Fort Collins, CO 80523 USA (e-mail: author@lamar. colostate.edu)}
% \address[Third]{Electrical Engineering Department, 
%    Seoul National University, Seoul, Korea, (e-mail: author@snu.ac.kr)}

\begin{abstract}                % Abstract of 50--100 words
The trade-off between model fidelity and computational cost remains a central challenge in the computational modeling of extrusion-based 3D printing, particularly for real-time optimization and control. Although high-fidelity simulations have advanced considerably for off-line analysis, dynamical modeling tailored for online, control-oriented applications is still significantly under-developed. In this study, we propose a reduced-order dynamical flow model that captures the transient behavior of extrusion-based 3D printing. The model is grounded in physics-based principles derived from the Navier–Stokes equations and further simplified through spatial averaging and input-dependent parameterization. To assess its performance, the model is identified via a nonlinear least-squares approach using Computational Fluid Dynamics (CFD) simulation data spanning a range of printing conditions, and subsequently validated across multiple combinations of training and testing scenarios. The results demonstrate strong agreement with the CFD data within the nozzle, the nozzle–substrate gap, and the deposited-layer regions. Overall, the proposed reduced-order model successfully captures the dominant flow dynamics of the process while maintaining a level of simplicity compatible with real-time control and optimization. 
\end{abstract}

% \begin{keyword}
% Five to ten keywords, preferably chosen from the IFAC keyword list.
% \end{keyword}

%\end{frontmatter}
%===============================================================================

\section{Introduction}

%\textcolor{red}{[Intro] Instruction: Here, write a few general statement about the 3D printing; why is it important and such; put some statistical numbers to signify its importance.}

\textcolor{black}{Extrusion-based additive manufacturing (AM), particularly direct ink writing (DIW), has emerged as a transformative approach in the fabrication of cementitious structures for civil infrastructure, architectural applications, and rapid construction \cite{mechtcherine2020extrusion}. This process involves the layer-by-layer deposition of a viscoplastic cement-based ink, offering geometric freedom and reduced material waste. Despite significant advancements in hardware and material formulations, the fundamental understanding of the real-time monitoring and control of the underlying flow dynamics remain limited, particularly under varying geometric, material, and operational conditions.}

%\textcolor{Now, write a small lit review on DIW of cementitious materials using FEM/FVM.} 

\textcolor{black}{Traditional modeling approaches, rooted in continuum mechanics and finite element methods, offer detailed insights into localized flow and stress distributions but are computationally expensive and sensitive to complex rheological behavior. Moreover, these models often require precise knowledge of boundary conditions and material properties, which may be difficult to obtain or change during the printing process.} These approaches convert partial differential equations (PDEs) to systems of algebraic equations that are solvable for computers. Examples of these approaches are the finite element method (FEM) and the finite volume method (FVM) \cite{roussel2020numerical}. There are multiple efforts in the existing literature that focus on modeling the flow of concrete using these simulation methods. 

A recent study focused on simulation-based flow modeling of five different non-Newtonian ink groups by numerically solving the conservation of mass and momentum equations \cite{sourov2025general}. The results reveal the effect of the inlet pressure on the printability of the inks. However, in this study, the printability of the cement-based ink and the effect of the input parameters on the printed geometry have not been discussed. Some studies in the literature specifically studied the geometric accuracy and cross-sectional shape of the printed material. For example, the cross-sectional shape and dimensions of the printed bead were analyzed by modeling the flow of the extruded material using FLOW-3D computational fluid dynamics (CFD) software \cite{rapha}. Different constitutive models were developed to study the shear-thinning and viscosity effects on the cross-section of the beads, and the results were compared with theoretical values used by the slicer software. To overcome the limitations of commercial software, including FLOW-3D, OpenFOAM framework was combined with the overInterDyMFoam solver to model 3D concrete printing to investigate the strand shape and cross-sectional geometry and dimensions in cases with various extrusion velocity, nozzle velocity and nozzle rotation \cite{wagner2025novel}. Similarly, an experimentally-validated CFD model was created to simulate strand cross-sections at different nozzle gap height and velocity ratios \cite{serdeczny2018experimental}. Despite the high resolutions of the aforementioned studies, there is a contradiction between resolution and efficiency in modeling of additive manufacturing of cementitious materials using the DIW process. Moreover, the models developed by these simulation methods are computationally heavy, making them unsuitable for real-time applications.

\textcolor{black}{In response to these challenges, data-driven dynamical modeling has emerged as a promising framework for characterizing and predicting the behavior of extrusion-based 3D printing processes. By leveraging high-resolution spatiotemporal data from experiments or simulations, these models can capture complex, nonlinear system dynamics with reduced-order representations, enabling real-time monitoring, control, and optimization.} The literature mostly focuses on the reduced-order control-oriented modeling of metal additive manufacturing techniques such as laser powder bed fusion \cite{wang2019control} and directed energy deposition \cite{wang2016reduced} to predict the geometrical characteristics of the printed part. However, there are only a few studies that focus on developing reduced-order models of the dynamics of the extrusion process. 

In this regard, in a recent study, thermo-viscoelastic behaviors of the filaments in the extrusion-based fused deposition modeling process were captured by a reduced-order model \cite{meng2024reduced}. Advanced models were also developed to understand spatial in-layer dynamics and layer-to-layer stability \cite{balta2019control}. In another study, the authors created a model of the filament behavior depending on the layer height \cite{percoco2020analytical}. The model was able to predict the required applied force at the inlet depending on the layer height. %\textcolor{red}{have you cited Prof. Qian Wang's papers?} \textcolor{gray}{cited in previous paragraph} 
Despite their efficiency and usefulness, these reduced-order models do not capture the evolution of the extruded strand as an explicit function of the governing process parameters. To the best of the authors’ knowledge, no existing study has formulated a reduced-order model specifically for the flow dynamics in DIW. This gap highlights the need for a computationally tractable yet physically meaningful model capable of predicting DIW strand formation based on controllable input variables.

%\textcolor{blue}{Now, write a small lit review on available reduced order model and their drawbacks}

\textcolor{black}{To address this research gap, we develop a physically-informed data-driven dynamical model in this study for the extrusion of cementitious materials, focusing on flow evolution along the material strand. The model integrates principles from physics-driven insights, reduced-order modeling, and system identification to capture the transient behavior of the ink-air interface during deposition. This modeling framework enables efficient prediction of strand quality and flow stability, providing a foundation for monitoring and closed-loop control in future smart manufacturing systems.}

\textcolor{black}{The structure of the paper is organized as follows. Section 2 provides a detailed overview of the extrusion-based 3D printing process and presents the derivation of a reduced-order dynamical model from a full-scale flow model incorporating the momentum and continuity equations. Section 3 describes the procedure for generating high-fidelity simulation data using ANSYS Fluent and explains how the parameters of the dynamical model are identified through least-squares fitting. Section 4 presents a comprehensive discussion and analysis of the results. Finally, Section 5 concludes the paper by summarizing the key findings and takeaways.}

% \section{Extrusion-based 3D Printing Process}

% \textcolor{red}{[Description] General description of the extrusion-based system. This section will be approx. 1 column}

% The schematic view of the direct ink writing (DIW) process is shown in Figure~\ref{fig:system1}. Printing steps include transporting build material by pumping (pushing material by an external force through the nozzle), deformation of build material during deposition (bending and extrudate swell), and the behavior of the build material after deposition on the moving plate, which is represented in our study by sub-system 1, sub-system 2, and sub-system 3, respectively. The material is pushed by a constant force in y direction resulting in a constant inlet mass flow rate to the nozzle. The nozzle is rigid and the build plate is moving with a constant velocity of $U_s$. The components of velocity in the x and y directions are $u$ and $v$, respectively. Detailed view of each sub-system, including the model variables is indicated in Figures  ~\ref{fig:subsystem1} and ~\ref{fig:subsystem3}. The inputs and outputs of each sub-system that will be used in the model derivation are shown in Figure ~\ref{fig:block}.

\section{Dynamical Modeling Framework for Extrusion-based 3D Printing Process}

In this section, we discuss the physical description of the extrusion-based 3D printing process, followed by the details of the proposed dynamical modeling framework.

\subsection{Physical description of the printing process}

%\textcolor{red}{[Description] General description of the extrusion-based system. This section will be approx. 1 column}

\begin{figure}[hbt!] % [h] places the figure approximately here
    \centering
    \includegraphics[width=0.7\linewidth]{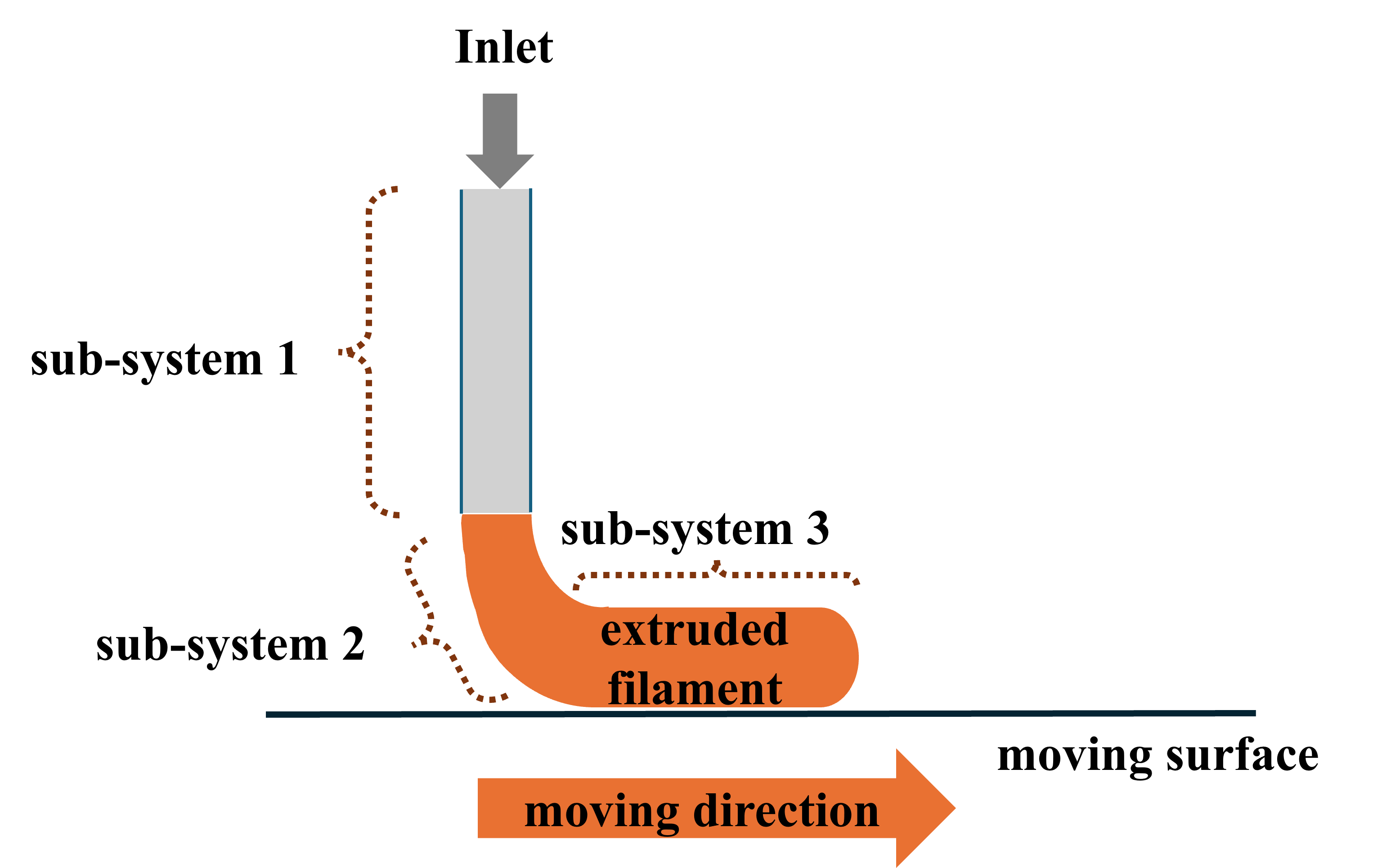} % Adjust the width as needed
    \caption{Schematic of the extrusion-based 3D printing process.}
    \label{fig:allsystem}
\end{figure}

The schematic of the DIW process is illustrated in Figure~\ref{fig:allsystem}. The printing operation can be decomposed into a sequence of interrelated physical processes: (i) material feeding, where the build material is introduced into the nozzle by an externally applied force (pumping), (ii) material deformation during deposition, encompassing phenomena such as bending, extrusion-induced swelling, and viscoelastic relaxation of the extrudate, and (iii) post-deposition dynamics, describing the evolution and stability of the deposited filament on the moving build plate. These stages are represented as sub-system 1, sub-system 2, and sub-system 3, respectively, in Figure~\ref{fig:allsystem}.

\begin{figure}[hbt!] % [h] places the figure approximately here
    \centering
    \includegraphics[width=0.7\linewidth]{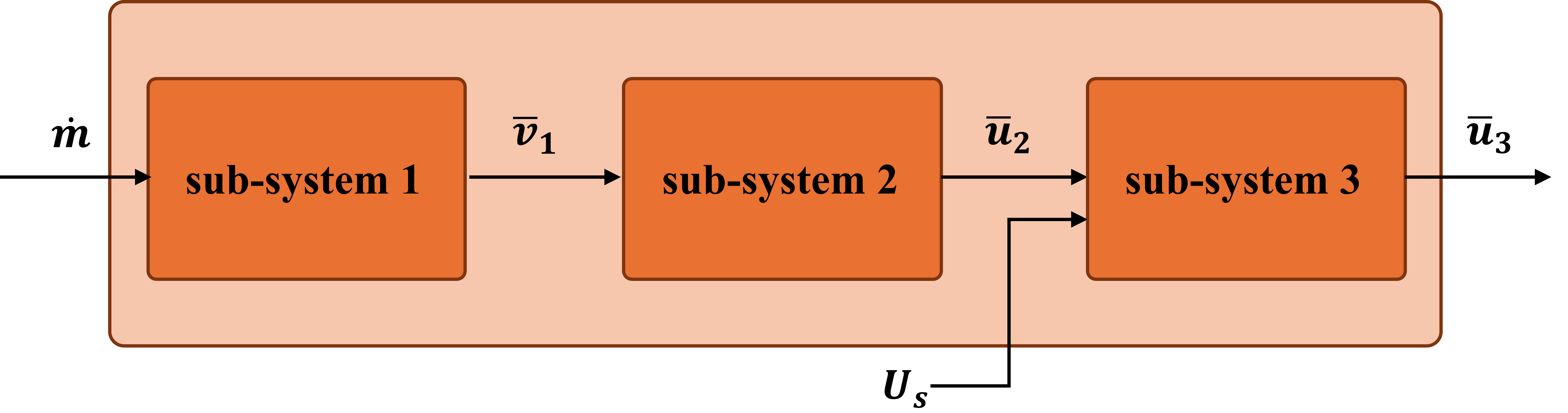} % Adjust the width as needed
    \caption{Block diagram representation of the input-output dynamics of the 3D printing process.}
    \label{fig:block}
\end{figure}

The corresponding inputs and outputs for each sub-system, which serve as the basis for the derivation of the dynamical model, are summarized in the block diagram shown in Figure~\ref{fig:block}. This decomposition enables a modular modeling approach, allowing each physical phenomenon to be represented with appropriate fidelity while maintaining computational tractability for system-level simulations.

\begin{figure}[hbt!] % [h] places the figure approximately here
    \centering
    \includegraphics[width=0.6\linewidth]{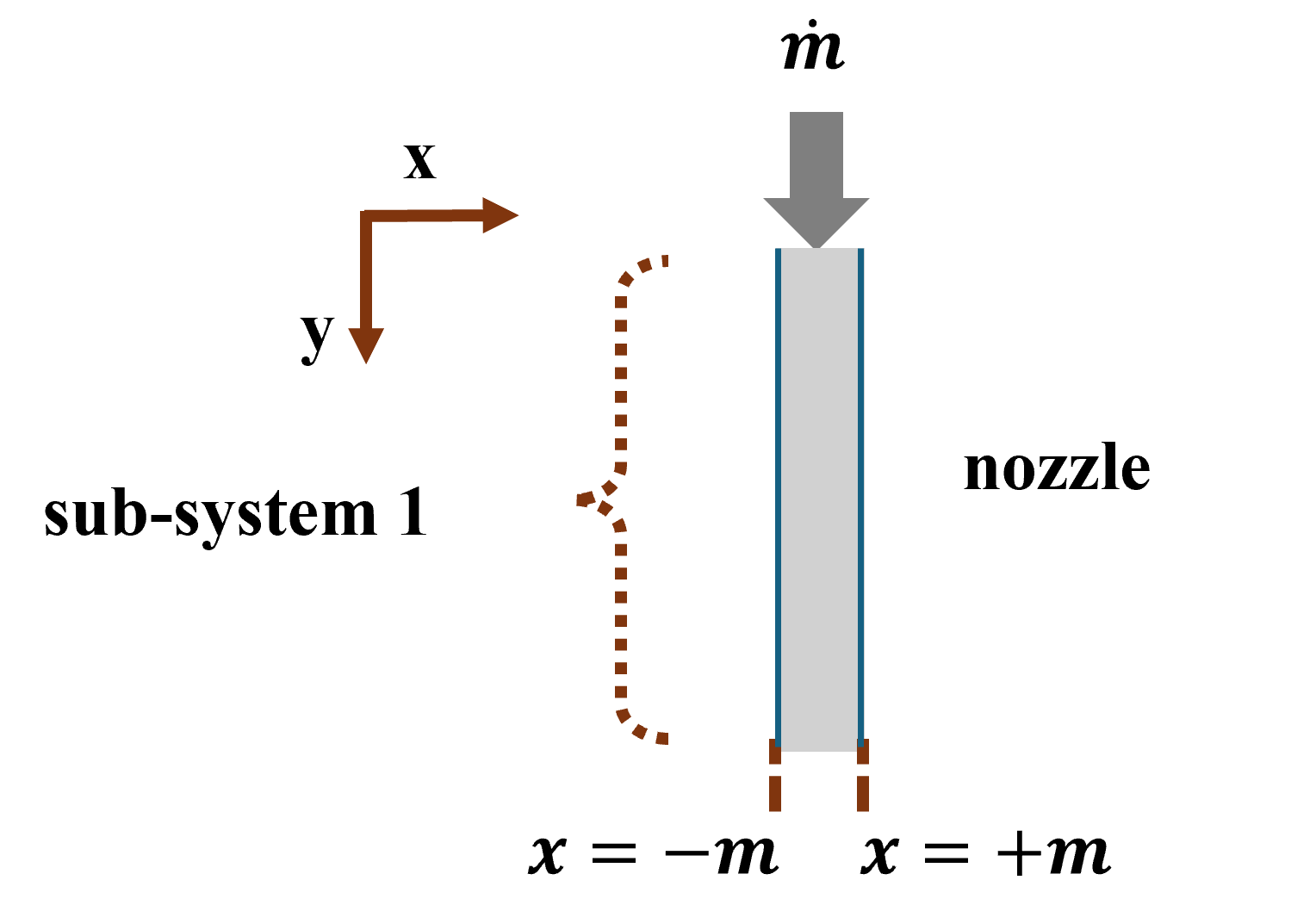} % Adjust the width as needed
    \caption{Schematic of sub-system 1.}
    \label{fig:subsystem1}
\end{figure}

In sub-system 1, the material is propelled by a constant force along the $y$-direction, producing a quasi-steady inlet mass flow rate at the nozzle. The nozzle is assumed rigid, while the build plate translates at a constant velocity $U_s$, imposing kinematic boundary conditions on the extrudate. As depicted in Figure~\ref{fig:subsystem1}, the velocity components of the material within the nozzle and during deposition are denoted as $u$ and $v$ in the $x$ and $y$ directions, respectively.

Figures~\ref{fig:subsystem1} and ~\ref{fig:subsystem3} provide detailed views of each sub-system, highlighting the key model variables required for reduced-order modeling. 

\subsection{Architecture of the dynamical model}

In this section, we discuss the details of the proposed modeling architecture. The physics of extrusion inside the nozzle involves complex dynamics due to the flow of a fluid with high viscosity, making it challenging to obtain extruded filaments with the desired shape and geometry. Furthermore, most modern control algorithms rely on a reasonably accurate system model which captures the essential physics. Despite the accuracy of the high-fidelity simulation models, they are not suitable for real-time control due to their high computational demand. Hence, to reduce the computational costs, we formulate a reduced order model which captures essential physical insights of the extrusion process -- without requiring significant computational overhead. 

The approach taken to achieve this goal is the following: We start with the physical laws, namely the continuity and momentum equations expressed as PDEs, which capture the physics dynamics of the process. Then, we incorporate spatial averaging technique on the spatio-temporal variables in momentum equations to get an averaged linear ordinary differential equation (ODE)-like mathematical structure. Next, we use the parameters of this approximate mathematical structure as tuning parameters of the model, and subsequently, apply nonlinear least squares based system identification technique to parameterize the model. Next, we discuss the details of this approach.

In order to enable a tractable and simplified modeling approach, we visualize the input-output dynamics of the overall printing process as a cascade connection of three sub-systems, as shown in Fig. \ref{fig:block}. The physical representations of these three sub-systems are shown in Fig. \ref{fig:allsystem}. First, we consider sub-system 1 where the input is the mass flow rate $\dot{m}$ of the material into the nozzle. The sub-system output is the flow velocity at the nozzle outlet. A simplified schematic of sub-system 1 is shown in Fig. \ref{fig:subsystem1}.

Considering the incompressible flow of a viscous fluid in a cylindrical vertical pipe without gravitational forces, the motion of the flow is governed by the continuity and momentum equations in a 2D cartesian coordinate system as follows:
\begin{align} 
    &\frac{\partial u_1}{\partial x}
+\frac{\partial v_1}{\partial y}
=0, \label{continuity}\\
   & \left(
\frac{\partial v_1}{\partial t}
+u_1\,\frac{\partial v_1}{\partial x}
+v_1\,\frac{\partial v_1}{\partial y}
\right)
= \nonumber\\
&-\frac{1}{\rho}\frac{\partial p_1}{\partial y}
+\nu\left(
\frac{\partial^2 v_1}{\partial x^2}
+\frac{\partial^2 v_1}{\partial y^2}
\right)+{f_t}_1, \label{y momentum}
\end{align}
where $p_1$ is the pressure, $\rho$ is the density, $\nu$ is the kinematic viscosity, and $u_1$ and $v_1$ are the x and y velocity components, respectively. Here, the effective force is ${f_t}_1$.

Furthermore, the velocity of the fluid is zero at the nozzle side walls, due to the no-slip boundary condition:

\begin{align} 
& v_1(x=-m,y)=0, \label{eq:bc1} \\
& v_1(x=m,y)=0,  \label{eq:bc2}
\end{align}

Finally, the initial velocity value is considered zero, i.e., $v_1(x,0)=0$, for all x. Next, we make the following assumptions to simplify these equations:

\begin{itemize}
\item \textbf{Assumption 1:} The fluid is treated as incompressible and Newtonian, and the flow is fully developed, implying spatial invariance of the velocity field in the streamwise direction of development.
\item \textbf{Assumption 2:} The flow possesses a single nonzero velocity component, such that \(u_1 = w_1 = 0\), and the motion is exclusively in the remaining coordinate direction.
\item \textbf{Assumption 3:} The flow is two-dimensional, varying only in the \(x\)- and \(y\)-directions, with no dependence on the \(z\)-coordinate.
\item \textbf{Assumption 4:} The flow regime is laminar, ensuring orderly, non-turbulent motion and smooth velocity gradients.
\item \textbf{Assumption 5:} There is no imposed pressure gradient in the \(x\)-direction, i.e., \(\partial p/\partial x = 0\).
\end{itemize}

Considering these assumptions, \eqref{continuity} and \eqref{y momentum} can be simplified as follows:
\begin{align}
&\frac{\partial v_1}{\partial y}=0, \label{y momentum222}\\
&\frac{\partial v_1}{\partial t} = -\frac{1}{\rho} \frac{\partial p_1}{\partial y} + \nu \frac{\partial^2 v_1}{\partial x^2} + {f_t}_1. \label{y momentum2}
\end{align}

Next, we proceed with spatio-temporal averaging of the variables in \eqref{y momentum222}-\eqref{y momentum2}. Due to the no-slip boundary condition on the nozzle walls, the velocity profile has a parabolic shape, which can be approximated by a second-degree polynomial with the equation of the following form:
\begin{equation} \label{eq:mean_velocity}
v_1(x,t)=a(t)(x^2-m^2).
\end{equation}
Hence, we apply the conservative averaging method (CAM), where the averaging of the velocity variable is performed as follows:
\begin{equation} \label{eq:avg_velocity}
\bar{v}_1 = \frac{1}{2m} \int_{-m}^{m} v_1(x,t) \,dx 
\end{equation}
Next, we differentiate \eqref{y momentum2} with respect to time $t$ to get
\begin{align}
    \frac{d\bar{v}_1}{dt}
& \nonumber= \frac{1}{2m}\int_{-m}^{m}\frac{\partial v_1}{\partial t}\,dx
= -\frac{1}{2m\rho}\int_{-m}^{m}\frac{\partial p_1}{\partial y}\,dx
\\&+ \frac{\nu}{2m}\int_{-m}^{m}\frac{\partial^{2}v_1}{\partial x^{2}}\,dx
+ \frac{1}{2m}\int_{-m}^{m} {f_t}_1\,dx . \label{eq-ss}
\end{align}
The individual terms on the right hand side of \eqref{eq-ss} are computed as shown below, using the expression \eqref{eq:mean_velocity}:
\begin{equation}
    -\frac{1}{2m\rho}\int_{-m}^{m}\frac{\partial p_1}{\partial y}dx=-\frac{1}{2m\rho}\frac{\partial p_1}{\partial y}\int_{-m}^{m}dx=-\frac{1}{\rho}\frac{\partial p_1}{\partial y}
\end{equation}
\begin{align} \label{inertia}
\frac{\nu}{2m}\int_{-m}^{m}\frac{\partial^{2}v_1}{\partial x^{2}}\,dx
& \nonumber= \frac{\nu}{2m}\left.\frac{\partial v_1}{\partial x}\right|_{-m}^{m}
= \frac{\nu}{2m}\left[\,2a x\,\right]_{-m}^{m}
\\&= \frac{\nu}{2m}\big(2am-(-2am)\big)
= 2a\nu .
\end{align}
Next, we find the relationship between the parameter $a$ and averaged velocity $\bar v_1$, as shown below:%Where $a$ here is obtained with respect to $\bar v$ as follows:
\begin{align}
\nonumber\bar{v}_1
\nonumber &= \frac{1}{2m}\int_{-m}^{m} a\,(x^{2}-m^{2})\,dx \\
\nonumber &= \frac{a}{2m}\left[\frac{x^{3}}{3}-m^{2}x\right]_{-m}^{m} \\
\nonumber &= \frac{a}{2m}\left(\frac{m^{3}}{3}-m^{3}-\left(-\frac{m^{3}}{3}+m^{3}\right)\right) \\
&= -\frac{2}{3}\,a\,m^{2} \nonumber \\
& \implies a=-\frac{3}{2}\frac{\bar v_1}{m^2}.\label{a definition}
\end{align}
% \begin{equation} \label{a definition}
%     a=-\frac{3}{2}\frac{\bar v}{m^2}
% \end{equation}
%By applying \eqref{eq:bc1} and \eqref{eq:bc2} boundary conditions to the equation \eqref{eq:mean_velocity} we have:
%\begin{equation}
%\bar{v} = \frac{-2}{3}am^2
%\end{equation}
%\begin{equation}
%am^2=-c
%\end{equation}

%Therefore, equation \eqref{eq:avg_velocity} becomes:
Inserting \eqref{a definition} into \eqref{inertia}, we have:
\begin{equation}
    \frac{\nu}{2m}\int_{-m}^{m}\frac{\partial^{2}v}{\partial x^{2}}\,dx=2a\nu=-3\nu\frac{\bar v_1}{m^2}
\end{equation}
%Since the velocity profile in the range of [-m,m] is continuous, the derivative of the average velocity with respect to time gives the following.

%\begin{equation}
%\frac{\partial \bar{v}}{\partial t} = \frac{1}{2m} \int_{-m}^{m} \frac{\partial v}{\partial t} \,dx 
%= \frac{1}{2m} \int_{-m}^{m} a \frac{\partial^2 v}{\partial x^2} \,dx 
%= 2 \alpha m a = -\frac{3a}{m} \bar{v}
%\end{equation}

Subsequently, we can write \eqref{eq-ss} as:
\begin{align}
    \frac{d\bar{v}_1}{dt}
= -\frac{1}{\rho}\frac{\partial p_1}{\partial y} -3\nu\frac{\bar v_1}{m^2}
+ \frac{1}{2m}\int_{-m}^{m} {f_t}_1\,dx . \label{eq-ss2}
\end{align}
Next, we utilize this physics-driven equation \eqref{eq-ss2} for average flow to formulate a reduced order flow model, which retains the structure of the equation but is parameterized differently. Note that the different parameterization is needed to ensure that we have enough tuning parameters in the model to fit it with the data. The new parameterized reduced order model is given by:
\begin{equation}
\frac{d\bar{v}_1}{dt}
= \beta_{1}\,{p_d}_1
+ \beta_{2}\,\bar{v}_1
+ \beta_{3}\,\dot{m},  \label{rom-1}
\end{equation}
where the first term $\beta_{1}\,{p_d}_1$ is inspired by the term $-\frac{1}{\rho}\frac{\partial p_1}{\partial y}$ in \eqref{eq-ss2} with ${p_d}_1$ being the pressure gradient, and the second term $\beta_{2}\,\bar{v}_1$ is inspired by the term $3\nu\frac{\bar v_1}{m^2}$ in \eqref{eq-ss2}. The third term $\beta_{3}\,\dot{m}$ is inspired by the term $\frac{1}{2m}\int_{-m}^{m} {f_t}_1\,dx$ -- where it is assumed that the major contribution of the force ${f_t}_1$ comes from the input mass flow rate $\dot{m}$. Here, the parameters $\beta_{1},\beta_{2}$ and $\beta_{3}$ are the tuning parameters of the model which need to be fitted to match the ground truth data. This concludes the reduced order modeling of sub-system 1, which captures the dynamic relationship between the inlet mass flow rate $\dot{m}$ and pressure gradient ${p_d}_1$ with the output average velocity $\bar{v}_1$.

% Furthermore, the pressure term of $\frac{\partial {p}}{\partial y}$ is a time-dependent function that can be written as $\beta_3(t)p_d$ due to the pressure drop over time.
% Therefore, according to the last equation, \eqref{y momentum2} can be reduced to an ODE in the following form.

% \begin{equation}
% \frac{d\bar{v}}{dt}
% = \beta_{1}(t)\,p_d
% + \beta_{2}\,\bar{v}
% + \beta_{3}\,\dot{m}
% \end{equation}
% e.g. one may take \beta_{2}=-3\nu/m^{2}.

% The sub-system 1 model inputs are the inlet mass flow rate $\dot m$ and the pressure gradient in $y$ direction ($p_d$), and the output is the average velocity of the fluid at the outlet of the nozzle. The input of the second sub-system is the output of the first sub-system and is equal to the average velocity of sub-system 1 ($\bar{v}_{i2}$). 
% $\beta_1(t)$, $\beta_2(t)$, and $\beta_3(t)$ are the model parameters. 

\begin{figure}[hbt!] % [h] places the figure approximately here
    \centering
    \includegraphics[width=0.6\linewidth]{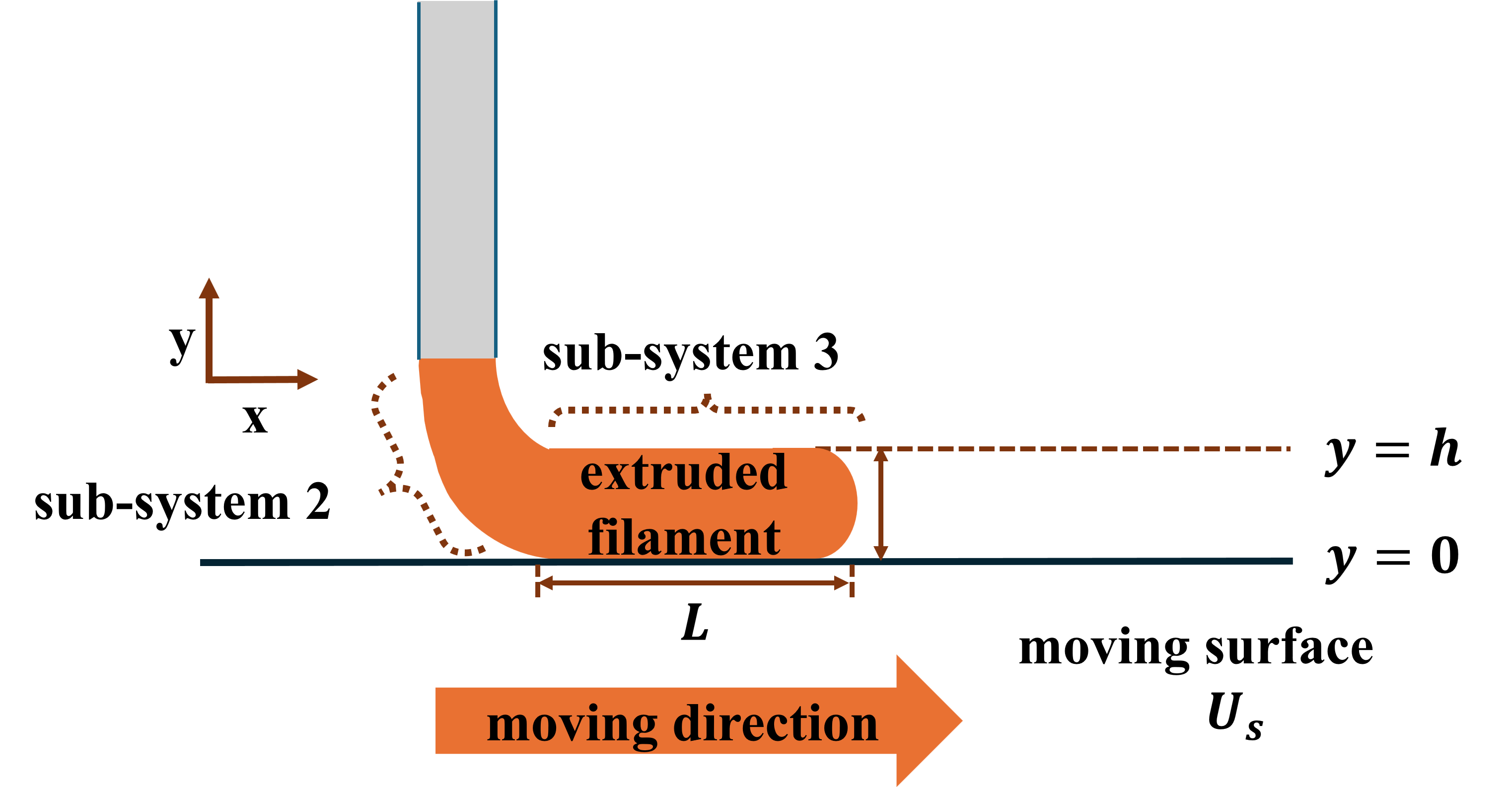} % Adjust the width as needed
    \caption{Schematic of sub-systems 2 and 3.}
    \label{fig:subsystem3}
\end{figure}

Next, we focus on sub-system 2 (refer to Fig. \ref{fig:subsystem3}) where the input is the sub-system 1 averaged velocity $\bar v_1$ and the output is the flow velocity at inlet of sub-system 3. The sub-system 2 is modeled by a linear algebraic mapping between the input and the output, as given below. 
\begin{equation}
    \bar u_{2}= \beta_4 \bar v_{1}, \label{rom-2}
\end{equation}
where $\bar u_2$ is the average velocity at the outlet of sub-system 2, and the parameter $\beta_{4}$ is the tuning parameter of the model which will be fitted to match the ground truth data later.

Now, we consider our last sub-system, namely sub-system 3 (refer to Fig. \ref{fig:subsystem3}) where the inputs are the sub-system 2 averaged velocity $\bar u_2$ and velocity of the moving surface $U_s$, and the output is the flow velocity at outlet of sub-system 3. We make the following assumptions:

\begin{itemize}
\item \textbf{Assumption 1:} The fluid is modeled as incompressible and Newtonian, implying constant density and a linear relationship between shear stress and strain rate.
\item \textbf{Assumption 2:} The flow field is two-dimensional, varying only in the \(x\)- and \(y\)-directions, with no dependence on the \(z\)-coordinate.
\item \textbf{Assumption 3:} The flow regime is laminar and fully developed, such that the velocity profile does not change in the streamwise direction.
\item \textbf{Assumption 4:} Body forces due to gravity are neglected, and thus gravitational acceleration does not enter the momentum balance.
\end{itemize}

The governing equations for sub-system 3 are similar to \eqref{continuity} and \eqref{y momentum}, with the following modifications: (i) The velocities are denoted as $u_3$ and $v_3$, instead of $u_1$ and $v_1$. (ii) The forcing term is dominated by the pressure gradient, which now depends on the sub-system 2 outlet velocity $\bar u_2$. 

Subsequently, the spatio-temporal flow velocity dynamics can be written as:
\begin{equation} \label{x momentum 3}
\frac{\partial u_3}{\partial t}
= -\frac{1}{\rho}\frac{\partial p_3}{\partial x}
+ \nu\,\frac{\partial^{2}u_3}{\partial y^{2}}.
\end{equation}

%Here, pressure gradient in x direction plays the roll of the source term to sub-system\#3. The inlet flow pressure to the sub-system\#3 ($p_{i3}(t)$) is achieved by reducing the pressure loss from the pressure delivered to the fluid by the applied force at the inlet of the nozzle and $f$ represent the approximate value of pressure gradient in principal direction of the flow. 

%\begin{equation} \label{x momentum 3}
%\frac{\partial u}{\partial t}
%= -\frac{1}{\rho}\frac{\partial p}{\partial x}
%+ \nu\,\frac{\partial^{2}u}{\partial y^{2}}
%- \underbrace{S_m(x,t)\,v_{\mathrm{i3}}(t)}_{S_u(x,t)} .
%\end{equation}

%Where $\bar{v}_{i3}$ is the average flow velocity at the inlet of the sub-system\#3.
The flow at $y=0$ is attached to the moving surface and due to the no-slip condition, the flow velocity at $y=0$ is equal to the velocity of moving surface, $U_s$. At $y=h$, the free-surface boundary condition is applied. The resulting boundary conditions are:
\begin{equation} \label{bc1}
    u_3(y=0)= U_s, \ 
\left.\frac{\partial u_3}{\partial y}\right|_{y=h}=0
\end{equation}
The governing equation \eqref{x momentum 3} along with the boundary conditions \eqref{bc1}, gives rise to a parabolic velocity profile (\cite{samanta2024combined}). Accordingly, we choose a second-degree polynomial to capture the velocity profile:
\begin{equation} \label{velocity profile3}
u_3(y,t)=a_3(t)y^2-b(t)y+c(t),
\end{equation}
Applying the boundary conditions \eqref{bc1} on \eqref{velocity profile3}, the coefficients $b$ and $c$ can be found, and the approximated velocity profile becomes as follows:
\begin{equation}
    c=U_s, \ b=-2a_3h \implies u_3=a_3 y^2 -2a_3 h y + U_s.
\end{equation}
Following the same CAM approach used for sub-system 1, we use an integral averaging of $u$ over $y$:
\begin{equation} \label{eq:avg_velocity}
\bar{u}_3 = \frac{1}{h} \int_{0}^{h} u_3(y,t) \,dy,
\end{equation}
which ultimately results in the following averaged velocity dynamics:
\begin{align} 
\frac{d\bar{u}_3}{dt}
&= \frac{1}{h}\int_{0}^{h}\frac{\partial u_3}{\partial t}\,dy
\nonumber \\& = -\frac{1}{h\rho}\int_{0}^{h}\frac{\partial p_3}{\partial x}\,dy
+ \frac{\nu}{h}\int_{0}^{h}\frac{\partial^{2}u_3}{\partial y^{2}}\,dy. \label{eq-39}
\end{align}
The first and second terms on the right hand side are computed as follows:
\begin{align}
    -\frac{1}{h\rho}\int_{0}^{h}\frac{\partial p_3}{\partial x}\,dy= -\frac{1}{h\rho}\frac{\partial p_3}{\partial x}\int_{0}^{h}dy=-\frac{1}{\rho}\frac{\partial p_3}{\partial x}, \label{eq-331}
\end{align}
\begin{align} \label{inertia 3}
\frac{\nu}{h}\int_{0}^{h}\frac{\partial^{2}u_3}{\partial y^{2}}\,dy
= \frac{\nu}{h}\left.\frac{\partial u_3}{\partial y}\right|_{0}^{h}
= \frac{\nu}{h}\left.\,(2a_3y-2a_3h)\,\right|_{0}^{h}
= 2a_3\nu .
\end{align}
Next, the parameter $a_3$ is obtained with respect to $\bar u_3$, as follows:
\begin{align}
\bar{u}_3 &\nonumber= \frac{1}{h}\int_{0}^{h} u_3(y,t)\,dy \\
& \nonumber= \frac{1}{h}\int_{0}^{h} \left(a_3 y^{2} - 2 a_3 h\, y + U_s\right)\,dy \\
&\nonumber= \frac{1}{h}\left[\frac{a_3 y^{3}}{3} - a_3 h y^{2} + U_s y\right]_{0}^{h} \\
&\nonumber= \frac{1}{h}\left(\frac{a_3 h^{3}}{3} - a_3 h^{3} + U_s h\right) \\
&= -\frac{2 a_3 h^{2}}{3} + U_s 
\Rightarrow a_3 = -\frac{3}{2h^{2}}\left(\bar{u_3} - U_s\right). \label{eq-222}
\end{align}

Considering \eqref{eq-331}, \eqref{inertia 3}, and \eqref{eq-222}, we can re-write \eqref{eq-39} as 
\begin{equation}
\frac{d\bar{u}_3}{dt}
= -\frac{1}{\rho}\,\frac{\partial p_3}{\partial x}
-\frac{3\nu}{h^2}\,\bar{u}_3\,+ \frac{3\nu}{h^2}\, U_s. \label{eq-ss29}
\end{equation}
Next, similar to sub-system 1, we utilize this physics-driven equation \eqref{eq-ss29} for average flow to formulate a reduced order flow model, which retains the structure of the equation but is parameterized differently. The new parameterized reduced order model is given by:
\begin{equation}
\frac{d\bar{u}_3}{dt}
= \beta_5 \bar u_2
+\beta_6\,\bar{u}_{3}\,+ \beta_7\, U_s, \label{rom-3}
\end{equation}
where the first term $\beta_{5}\,\bar u_2$ is inspired by the term $-\frac{1}{\rho}\frac{\partial p_3}{\partial x}$ in \eqref{eq-ss29} with $p_3$ assumed to be mainly driven by sub-system 2 flow velocity $\bar u_2$, and the second term $\beta_{6}\,\bar{u}_{3}$ is inspired by the term $-\frac{3\nu}{h^2}\,\bar{u}$ in \eqref{eq-ss29}. The third term $\beta_7\, U_s$ is inspired by the term $\frac{3\nu}{h^2}\, U_s$ in \eqref{eq-ss29}. Here, the parameters $\beta_{5},\beta_{6}$ and $\beta_{7}$ are the tuning parameters of the model which will be fitted to match the ground truth data later. This concludes the reduced order modeling of sub-system 3, which captures the dynamic relationship between the inlet flow velocity $\bar u_2$, moving surface velocity $U_s$, and the output average velocity $\bar{u}_3$.

\subsection{Input-dependent model parameterization}

In the previous subsection, we developed a dynamical model consisting of three coupled differential-algebraic equations \eqref{rom-1}, \eqref{rom-2}, and \eqref{rom-3}. Note that this dynamical model is linear in terms of input and output. This simplified structure of the model will be useful for developing closed-form feedback control solutions using model-based analytical techniques. However, this linear structure sacrifices accuracy for the sake of simplicity. In order to enhance its accuracy, we further parameterize the model coefficients ($\beta$'s in \eqref{rom-1}, \eqref{rom-2}, and \eqref{rom-3}). Note that the main inputs to the overall system are the inlet mass flow rate $\dot{m}$ and the velocity of the moving plate $U_s$. Depending on various inlet mass flow rates and plate velocities the underlying physical process is expected to show different dynamics, owing to the nonlinearities in the process. In order to capture such input dependencies, we propose the following parameterization of the model coefficients:
\begin{align}
    & \beta_1 = \gamma_1 \dot{m}, \ \beta_2 = \gamma_2 \dot{m}, \ \beta_3 = \gamma_3 \dot{m}, \label{param-1}\\
    & \beta_5 = \gamma_5 U_s, \ \beta_6 = \gamma_6 U_s, \ \beta_7 = \gamma_7 U_s. \label{param-2}
\end{align}

Here, we assumed that the velocity dynamics of sub-system 1 depend on the inlet mass flow rate, since this sub-system is directly affected by material injection at the inlet, and the underlying nonlinearities are captured by \eqref{param-1}, where the model coefficients are expressed as a linear function of $\dot{m}$. Similarly, the velocity dynamics of sub-system 3 depend on the plate velocity since this sub-system is in direct contact with the plate, and the underlying nonlinearities are captured by \eqref{param-2}, where the model coefficients are expressed as a linear function of $U_s$.

%\begin{equation} \label{eq:avg_velocity x}
%\bar{u} = \frac{1}{h} \int_{0}^{h} u \,dy = \frac{1}{h} \int_{0}^{h} (dy^2 + ey + f) \,dy = \frac{dh^2}{3} +\frac{eh}{2} + f
%\end{equation}

%\begin{equation}
%\frac{\partial \bar{u}}{\partial t} = \frac{1}{h} \int_{0}^{h} \frac{\partial u}{\partial t} \,dy 
%= \frac{1}{2m} \int_{-m}^{m} a \frac{\partial^2 v}{\partial x^2} \,dx 
%= 2 \alpha m a = -\frac{3a}{m} \bar{v}
%\end{equation}

%\begin{equation} \label{eq:ray}
%\rho(\frac{\partial v}{\partial t}+U\frac{\partial u}{\partial x}+V\frac{\partial v}{\partial y} = - \frac{\partial p}{\partial x} + \eta (\frac{\partial^2 u}{\partial x^2}+\frac{\partial^2 v}{\partial y^2}) + \rho g_x
%\end{equation}
%Considering that the flow only exists due to the motion of the plate and there is no pressure gradient, and no slip condition on the plate, the boundary and initial conditions are:
%\begin{equation}
%v(0)=u_{22}
%\end{equation}
%$u_{22}$ is defined as the output velocity of sub-system 2.
%\begin{equation}
%v(0,t)=u_p
%\end{equation}
%\begin{equation}
%v(\infty,t)=0
%\end{equation}
%Which means no flow exists far from the plate.
%\begin{equation}
%\frac{\partial u}{\partial x}=\frac{\partial^2 u}{\partial x^2}=0
%\end{equation}
%The effect of gravity on the flow motion is neglected. 
%\begin{equation}
%g_x=0
%\end{equation}
%\begin{equation}
%\frac{\partial p}{\partial x}=0
%\end{equation}
%Therefore equation \eqref{eq:ray} reduces to the following equation:
%\begin{equation}
%\frac{\partial u}{\partial t}=\nu\frac{\partial^2 u}{\partial y^2}
%\end{equation}

\section{Data Generation and Reduced-Order Model Fitting}

\textcolor{black}{In this section, we first describe the methodology used to generate data from the high-fidelity computational model. We then outline the procedures employed for training and testing the dynamical model.}

\subsection{Data generation from high-fidelity computational model}

%\textcolor{red}{check grammar of Section 3.1 and update - some are in past and some are in present}

\textcolor{black}{High-fidelity simulation models are commonly employed to evaluate the performance of the proposed model. CFD enables the numerical solution and analysis of the Navier–Stokes equations without requiring major simplifying assumptions. In this study, velocity and pressure fields within the three subsystems of the extrusion process are obtained from a 3D CFD simulation executed on an Intel Xeon\textsuperscript{®} Gold 6230R CPU @ 2.10 GHz workstation with 128 GB of RAM. ANSYS Fluent (Workbench 2022R1) is used to simulate the flow of a cement-based material through the extrusion nozzle and its subsequent deposition onto a build plate. A pressure-based, transient solver with absolute velocity formulation is employed. Pressure–velocity coupling is achieved using the PISO algorithm, with both skewness and neighbor correction parameters set to 1. The PRESTO scheme is applied for pressure interpolation, second-order upwind for momentum, and compressive for volume-fraction interpolation. The least-squares cell-based method is used for gradient reconstruction, and a relaxation factor of 0.75 is applied to the flow variables.} \textcolor{black}{The CFD model has been validated against available literature results \cite{liu2020modelling}, showing good agreement in deposition profile. A detailed discussion of the validation and comparison with experimental results will be presented in a forthcoming publication.}

\textcolor{black}{For the multiphase flow model in ANSYS Fluent, the properties of two phases are specified: the cement-based material being extruded and the air occupying the nozzle and deposition region prior to flow initiation. Interactions between the phases are defined using an implicit formulation of the volume-fraction equations, with a cutoff value of 1e-6. The interface-modeling approach is set to sharp/dispersed, and implicit body-force formulation is enabled. Surface-tension effects are included using the continuum surface stress model, with a surface-tension coefficient of 0.94 N/m between the two phases.}

\begin{figure}[hbt!] % [h] places the figure approximately here
    \centering
    \includegraphics[width=0.9\textwidth]{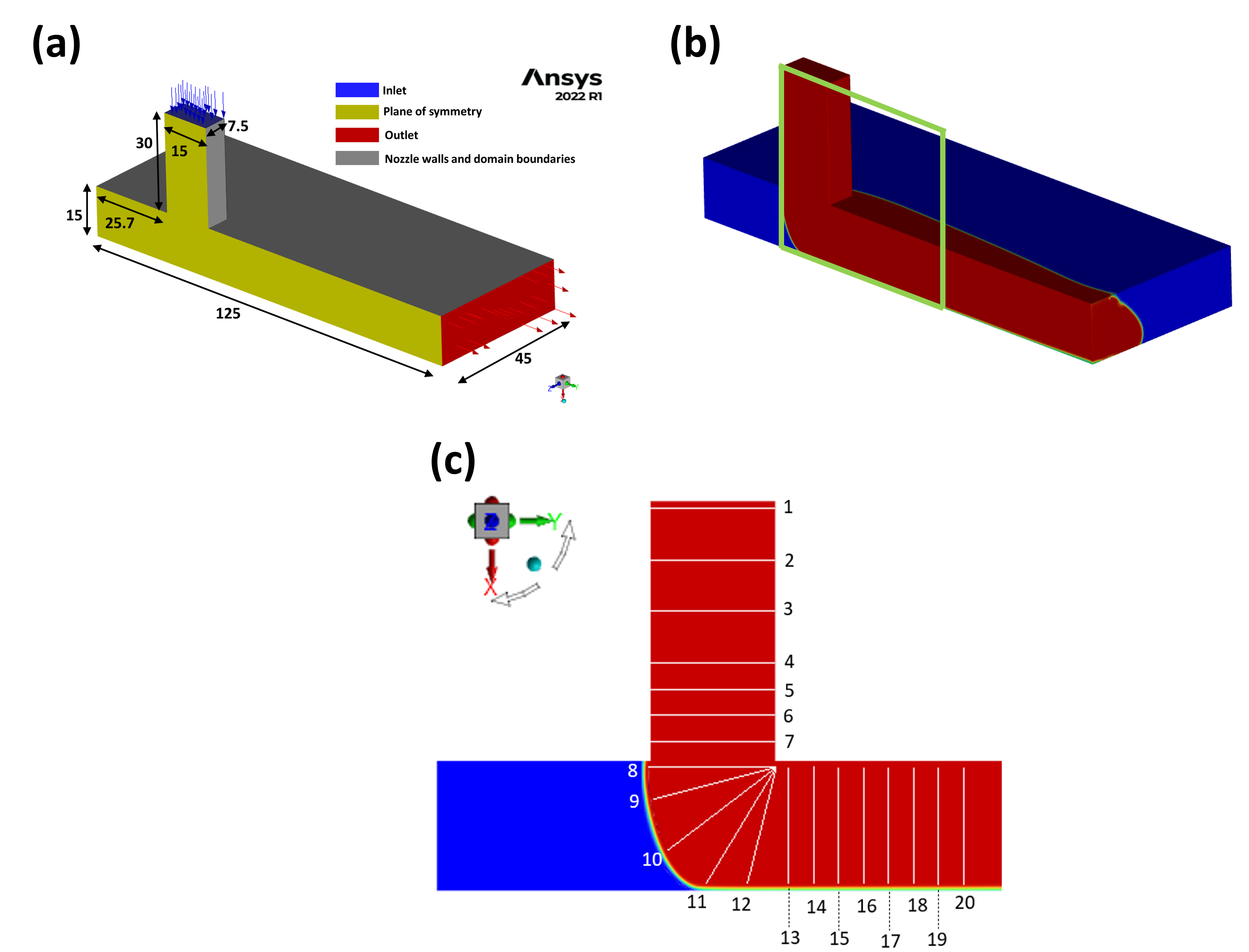} % Adjust the width as needed
    \caption{(a) High-fidelity computational domain. All dimensions are in mm. (b) A representative DIW strand. The green box indicates the cross-sectional area for (c). (c) Cross-sectional locations of the DIW strand from where the velocity and pressure data are extracted.}
    \label{fig:cfd}
\end{figure}

\textcolor{black}{The computational domain is shown in \ref{fig:cfd}(a). The nozzle length is 30 mm, and the exit plane is positioned 15 mm above the build plate. The build plate is modeled as a moving wall to represent the relative lateral motion between the nozzle and substrate during deposition. All walls are assigned no-slip boundary conditions. Assuming axisymmetric flow, only half of the domain is modeled, and a symmetry condition is applied along the nozzle centerline plane.}

\textcolor{black}{A mesh-independence analysis is conducted to ensure that the mesh resolution is sufficient to accurately capture the strand shape and the phase-interface quality. A hex-dominant mesh with a characteristic size of 3e-4 m is used. The CFD-generated training dataset consists of nine case studies spanning a range of surface velocities and inlet mass-flow rates. These cases are summarized in Table \ref{tab:sim_cases}.} \textcolor{black}{A representative strand is shown in Fig.~\ref{fig:cfd}(b). In this figure, the red regions denote the cementitious matrix, while the blue regions indicate air-filled zones. The green box marks the cross-sectional area corresponding to Fig.~\ref{fig:cfd}(c), where the positions of the cross-sections are illustrated. For each simulation, time-resolved velocity and pressure fields are obtained by computing cross-sectional averages of these quantities at predefined locations within the flow domain. Cross-sections 1–7 correspond to sub-system 1, 8–12 correspond to sub-system 2, and 13–20 correspond to sub-system 3.}

%\begin{table}[h]
%\centering
%\begin{tabular}{|>{\centering\arraybackslash}m{4.2cm}|*{5}{>{\centering\arraybackslash}m{1.8cm}|}}
%\hline
%\diagbox[width=4.2cm,height=1.8cm]{\textbf{Substrate}\\ \textbf{Speed}\\(mm/s)}%
%{\textbf{Inlet}\\ \textbf{Flow Velocity}\\(mm/s)}
 % & \textbf{50} & \textbf{55} & \textbf{60} & \textbf{65} & \textbf{70} \\
%\hline
%\textbf{50} & X &   & X &   & X \\
%\hline
%\textbf{55} &   &   &   &   &   \\
%\hline
%\textbf{60} & X &   & X &   & X \\
%\hline
%\textbf{65} &   &   &   &   &   \\
%\hline
%\textbf{70} & X &   &   &   & X \\
%\hline
%\end{tabular}
%\caption{Training and testing datasets.}
%\end{table}

\subsection{Training and testing of the proposed dynamical model}

In this subsection, we utilize the data generated from simulation cases in Table \ref{tab:sim_cases} to train the proposed dynamical model (given by \eqref{rom-1}, \eqref{rom-2}, and \eqref{rom-3}) and test its performance. The ultimate objective of model training is to tune the model parameters (i.e., $\beta_4$ in \eqref{rom-2} and $\gamma$'s in \eqref{param-1} and \eqref{param-2}) to minimize the loss between the model's output and the ground-truth data generated by the CFD model. We have utilized a nonlinear least-squares technique to fit the model parameters in MATLAB  \cite{dennis1977nonlinear}. In the nonlinear least-squares setting, the following objective function is minimized:
\begin{equation}
    \min_{x} f(x) = \sum_{i=1}^{n} f_i(x)^2,
\end{equation}
where $f_i(x)$ are auxiliary functions, and the algorithm minimizes the sum of their squares. Suppose that the observed-data from the CFD simulation is in the form of $\{(t_i, y_i)\}_{i=1}^n$ and the model is in the form of $\hat{M}(\gamma)$, then the algorithm solves the following equation to estimate parameter $\gamma$:
\begin{equation}
    \min_{\gamma} f(\gamma)
= \sum_{i=1}^{n} \bigl(\hat{M}(\gamma) - y_i\bigr)^2
\end{equation}

Hence, the model $\hat{M}$ is essentially a time-discretized version of the model \eqref{rom-1}, \eqref{rom-2}, and \eqref{rom-3} (using Euler's first order discretization) and the parameters are $\beta_4$ and $\gamma_j$, where $j \in \{1,2,3,5,6,7\}$.

\begin{table}[hbt!]
  \centering
  \caption{Simulation cases for model training and testing.}
  \label{tab:sim_cases}
  \setlength{\tabcolsep}{3pt}   % <-- smaller horizontal padding
  \small                        % <-- slightly smaller font
  \begin{tabular}{cccc}
    \toprule
    Case \# & Inlet flow       & Inlet mass flow & Surface \\
          & velocity [mm/s]  & rate ($\dot{m}$) [kg/s]     & velocity ($U_s$) [mm/s] \\
    \midrule
    1 & 50 & 0.02475   & 50 \\
    2 & 50 & 0.02475   & 60 \\
    3 & 50 & 0.02475 & 70 \\
    4 & 60 & 0.02970  & 50 \\
    5 & 60 & 0.02970 & 60 \\
    6 & 60 & 0.02970 & 70 \\
    7 & 70 & 0.03465 & 50 \\
    8 & 70 & 0.03465 & 60 \\
    9 & 70 & 0.03465 & 70 \\
    \bottomrule
  \end{tabular}
\end{table}

\textcolor{black}{In this study, the available CFD-generated simulation cases are partitioned into training and testing subsets to support the development and evaluation of the proposed dynamical model. The training set is selected to provide broad coverage of the range of inlet mass flow rates and surface velocities represented in the simulations, thereby ensuring that the model is exposed to a diverse set of flow conditions during parameter identification. The testing set is reserved exclusively for performance evaluation and includes cases not used during training, enabling an unbiased assessment of the model’s generalization capability.}

\section{Results and Discussion}

\subsection{High-fidelity data analysis}

\textcolor{black}{The high-fidelity CFD simulations generate a systematically curated dataset that characterizes the multiphase flow behavior of the cementitious material under controlled variations in inlet and surface operating conditions. In total, nine simulation cases are produced, each corresponding to a unique combination of inlet mass flow rate and surface (plate) velocity. These parameters are selected to span the representative operating envelope of the DIW extrusion process and to capture the resulting variations in flow kinematics, pressure distribution, and shear-induced deformation within the material.}

\textcolor{black}{The inlet velocity is prescribed at three discrete values, 50 mm/s, 60 mm/s, and 70 mm/s, corresponding to inlet mass flow rates of 0.02475 kg/s, 0.02970 kg/s, and 0.03465 kg/s, respectively.  Independently, the surface velocity is varied across the same three levels (50 mm/s, 60 mm/s, 70 mm/s) to emulate different nozzle–substrate relative motion speeds during deposition. These values are selected based on the prescribed values in the literature for the DIW of the cement-based materials \cite{liu2020modelling}. A full factorial combination of these two parameters results in the nine simulation cases summarized in Table \ref{tab:sim_cases}, ensuring that both primary process inputs are uniformly sampled across the design space.}

\textcolor{black}{For each simulation, time-resolved pressure and velocity fields are extracted and subsequently reduced to cross-sectional averages at multiple predefined locations along the extrusion flow path. These reduced-order quantities serve as meaningful descriptors of the underlying flow physics, preserving the essential spatial trends and sensitivities of the system while filtering out small-scale fluctuations not critical to model development.}

\begin{figure}[hbt!] % [h] places the figure approximately here
    \centering
    \includegraphics[width=0.95\textwidth]{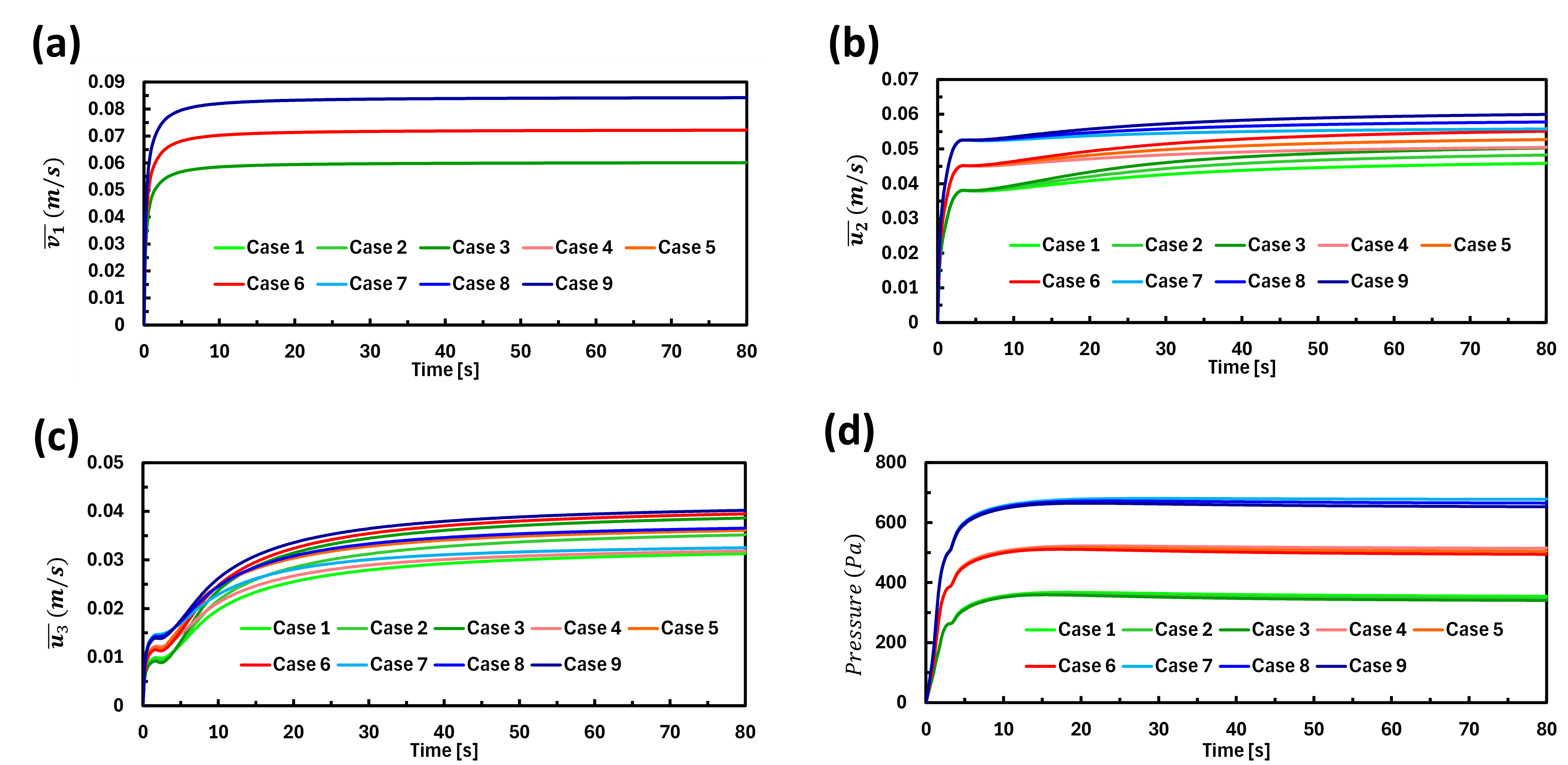} % Adjust the width as needed
    \caption{Average velocity profiles for a) sub-system 1, b) sub-system 2, c) sub-system 3. d) Average pressure profile for sub-system 1.}
    \label{fig:hf_data}
\end{figure}

\textcolor{black}{Analysis of the full simulation dataset for sub-system 1 (Figure \ref{fig:hf_data}(a)) reveals clear and physically consistent trends. Increasing the inlet mass flow rate produces proportional increases in both the average axial velocity and static pressure within the nozzle interior, reflecting the expected momentum input imposed by the extrusion system. In contrast, variations in surface velocity exhibit no measurable influence on the flow within this region. The corresponding datasets are nearly indistinguishable: Simulations 2–3, 5–6, and 8–9 fully overlap with simulations 1, 4, and 7, respectively. These results confirm that the flow behavior in sub-system 1 is governed exclusively by the inlet mass flow rate, with the plate motion exerting negligible upstream influence due to the dominance of pressure-driven flow inside the nozzle.}

\textcolor{black}{In sub-system 2 ((Figure \ref{fig:hf_data}(b)), the combined influence of the inlet flow and surface velocities becomes more pronounced. Here, the effect of plate motion is non-negligible: as the surface velocity increases, the average material velocity increases as well. This behavior arises from an entrainment mechanism in which the moving substrate induces additional shear at the interface, partially dragging the extrudate in the direction of motion. As a result, the flow in this transitional region reflects a coupled response to both the imposed inlet momentum and the substrate-driven deformation.}

\textcolor{black}{Sub-system 3 ((Figure \ref{fig:hf_data}(c)) exhibits a markedly different sensitivity profile. Once the material has exited the nozzle and fully engages with the moving build plate, the influence of inlet mass flow rate diminishes substantially. Instead, the velocity field becomes governed almost entirely by the plate velocity. In this region, the extrudate behaves as a transported body whose downstream kinematics are dictated primarily by substrate motion rather than upstream extrusion dynamics.}

\textcolor{black}{Figure \ref{fig:hf_data}(d) illustrates the static pressure distribution within sub-system 1 and reinforces the conclusions drawn from the velocity analysis in Figure \ref{fig:hf_data}(a). As the inlet mass flow rate increases, the pressure throughout the nozzle rises proportionally, consistent with the expected behavior of a pressure-driven internal flow. Variations in surface velocity do not introduce any measurable deviation in the pressure field, further confirming that plate motion has no upstream influence within this region. The close correspondence between the pressure profiles across simulations with identical mass flow rates, regardless of surface velocity, demonstrates that the hydrodynamic state of sub-system 1 is controlled solely by the imposed inlet conditions.}

\textcolor{black}{Taken together, these observations provide a consistent and mechanistic understanding of flow behavior across the DIW process. The nozzle interior (sub-system 1) operates in a pressure-dominated regime, where the inlet mass flow rate is the primary driver. The downstream region on the build plate (sub-system 3) operates in a kinematically driven regime, where substrate motion dominates. The intermediate region (sub-system 2) embodies a coupled regime in which both inputs shape the velocity field, reflecting the interplay between extrusion-induced flow acceleration, strand formation, and substrate-imposed shear. This physics-based interpretation forms the foundation for subsequent model training and validation across varying process conditions.}

\subsection{Model evaluation for different mass flow conditions}

%\textcolor{red}{training data: Case 1, 2, 3, 7, 8, 9}

%\textcolor{red}{This section will contain three figures. The first figure will include 3 columns $\times 9$ rows of subfigures. Each row will represent one of the 9 cases. Each column will represent 3 sub-systems. EXPLAIN RESULTS THE WAY IT IS DONE FOR SECTION 4.1}

%\textcolor{red}{The second figure will have the RMSE distribution of the dynamical model like Fig. 15. The third figure will have the probability distribution like Fig. 16}

\textcolor{black}{In this subsection, we first assess the ability of the dynamical model to perform interpolation with respect to the inlet mass flow rate. To this end, the model is trained using datasets corresponding to the lowest and highest mass flow rates in the simulation set, specifically 0.02475 kg/s and 0.03465 kg/s.} \textcolor{black}{After training, the model is evaluated on the intermediate mass flow rate of 0.02970 kg/s, a condition not included in the training set. This testing configuration enables investigation of the model’s capability to generalize and accurately predict flow dynamics for mass flow rates that lie between the trained extremes.}

\begin{figure}[hbt!] % [h] places the figure approximately here
    \centering
    \includegraphics[width=0.95\textwidth]{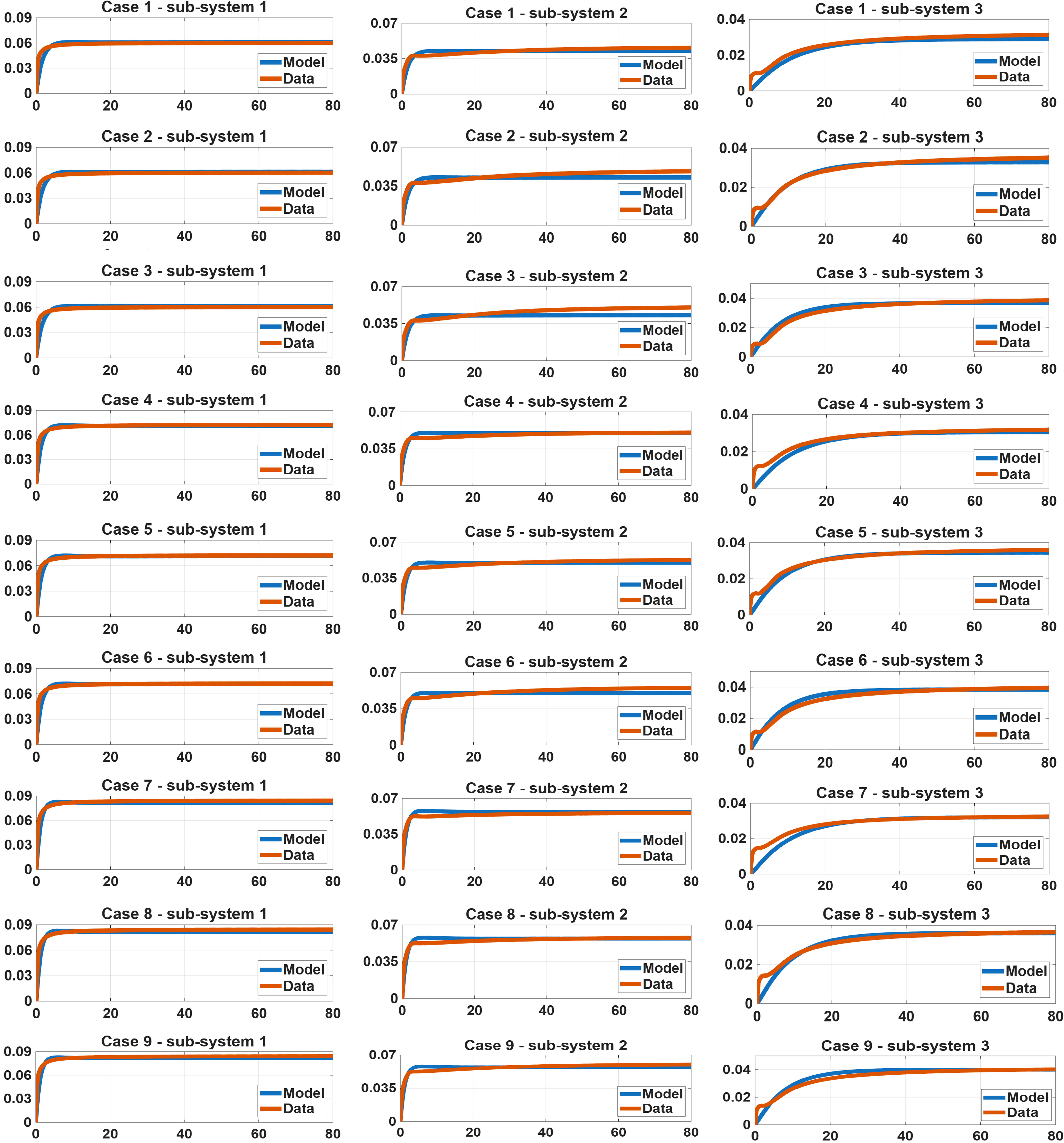} % Adjust the width as needed
    \caption{Performance of the dynamical model trained with simulation cases 1, 2, 3, 7, 8, and 9. The model is evaluated on intermediate flow conditions (cases 4, 5, and 6) not used during training. ``Data'' indicates CFD simulation data and ``Model'' indicates proposed dynamical model. The $y$-axis for the first column corresponds to $\bar{v}_1$ (mm/s), the second column to $\bar{u}_2$ (mm/s), and the third column to $\bar{u}_3$ (mm/s), while the $x$-axis for all figures represents time in seconds (s).}
    \label{fig:int_mfr_all}
\end{figure}

\textcolor{black}{In this interpolative scenario, the simulation cases are organized according to the inlet mass flow rate, ranging from the lowest (Cases 1–3) to the highest (Cases 7–9), while the intermediate mass flow rates (Cases 4–6) are reserved for testing the predictive capabilities of the model. This setup allows us to evaluate the model’s ability to interpolate between training points rather than simply reproducing previously seen conditions. Figure \ref{fig:int_mfr_all} presents the model predictions for each sub-system across all simulation cases. The $y$-axis for the first column corresponds to $\bar{v}_1$ (mm/s), the second column to $\bar{u}_2$ (mm/s), and the third column to $\bar{u}_3$ (mm/s), while the $x$-axis for all figures represents time in seconds (s). The results indicate that the predicted velocity fields closely replicate the corresponding ground truth data for both the low- and high-mass flow regimes, as well as the interpolated intermediate cases. Across sub-systems 1–3, the model captures not only the overall trends but also the subtle variations in velocity induced by changes in inlet mass flow and surface velocities. These observations confirm that the model generalizes well within the range of training data, accurately representing the underlying physical behavior even for cases it has not encountered during training.}

%\textcolor{red}{training data, extrapolation 1: Case 4, 5, 6, 7, 8, 9}

%\textcolor{red}{training data, extrapolation 2: Case 1, 2, 3, 4, 5, 6}

%\textcolor{red}{This section will contain 4 figures: RMSE for extrapolation 1, probability distribution for extrapolation 1, RMSE for extrapolation 2, and probability density for extrapolation 2}

\begin{figure}[hbt!] % [h] places the figure approximately here
    \centering
    \includegraphics[width=0.6\textwidth]{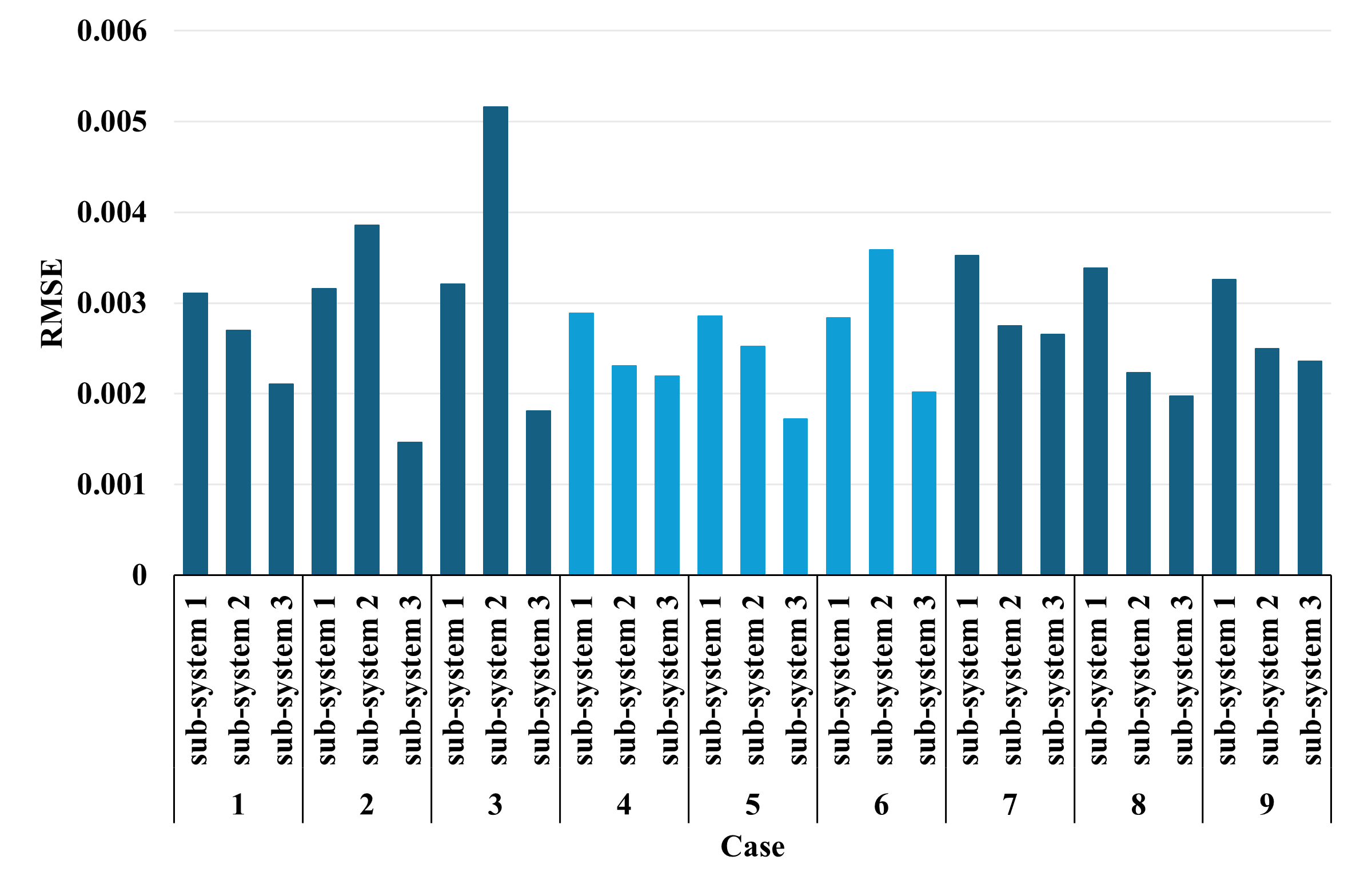} % Adjust the width as needed
    \caption{RMSE distribution obtained from the dynamical model across all sub-systems for training and testing datasets in the interpolative mass flow rate scenario. The dynamical model is trained with cases 1, 2, 3, 7, 8, and 9 and tested on cases 4, 5, and 6. Dark blue denotes the training cases, whereas light blue denotes the testing cases.}
    \label{fig:Int_MFR_RMSE}
\end{figure}

\textcolor{black}{The root mean squared error (RMSE) distribution (Figure \ref{fig:Int_MFR_RMSE}) reveals distinct behaviors across the three sub-systems. For sub-system 1, the RMSE values for the testing cases are lower than those for the training cases, indicating that the model, trained on the extreme mass flow rates, can predict intermediate flow conditions with even higher accuracy.}

For sub-system 2, the RMSE for the testing cases falls between the training errors. This behavior reflects the combined influence of inlet mass flow and surface velocity in this transitional region: since the model has been trained only on the extreme mass flow conditions, it must interpolate both mass flow and substrate-driven effects. The intermediate RMSE values suggest that the model successfully captures the coupled response, though with slightly higher uncertainty than in sub-system 1.

{Similarly, for sub-system 3, the testing RMSE also lies between the training errors. In this region, the velocity field is dominated primarily by the plate motion rather than the inlet flow, and the model’s ability to generalize from the training cases results in moderate prediction errors for the intermediate mass flow conditions. Overall, the RMSE trends across all sub-systems indicate that the model exhibits robust interpolative capability, with prediction accuracy influenced by the relative dominance of flow-driving mechanisms in each sub-system.}

\begin{figure}[hbt!] % [h] places the figure approximately here
    \centering
    \includegraphics[width=0.6\textwidth]{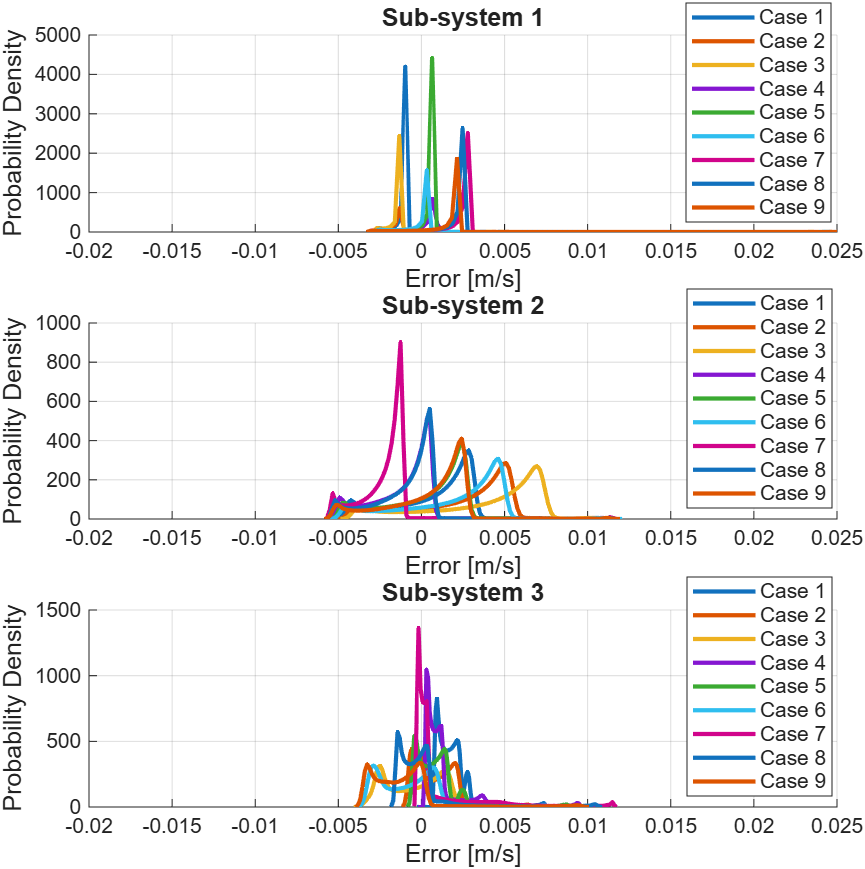} % Adjust the width as needed
    \caption{Probability distribution of model errors for the interpolative mass flow rate training scenario. The dynamical model is trained with cases 1, 2, 3, 7, 8, and 9 and tested on cases 4, 5, and 6.}
    \label{fig:Int_MFR_ED}
\end{figure}

The probability distributions of the prediction errors (Figure \ref{fig:Int_MFR_ED}) demonstrate distinct performance characteristics across the three sub-systems. For sub-system 1, the error distribution is sharply peaked around zero, indicating that the model exhibits high fidelity in reproducing the true velocity field. The testing cases—corresponding to intermediate mass-flow-rate conditions not explicitly included in the training set—show particularly narrow distributions. This behavior reflects strong interpolative generalization, suggesting that the model successfully captures the underlying physical mapping between mass flow rate and velocity without overfitting to the boundary conditions used during training. Across all operating conditions, the errors remain confined within approximately $-0.005$ m/s to $0.005$ m/s, underscoring the model’s excellent predictive accuracy and numerical stability.

For sub-system 2, the error distributions span a wider range, as illustrated in Figure \ref{fig:Int_MFR_ED}. This broader spread suggests increased sensitivity of the learned mapping to geometric or flow-regime variations present in this part of the system. Nonetheless, the error envelopes of the testing cases remain largely contained within those associated with the training cases. This indicates that, despite the higher variability, the model preserves reliable generalization for intermediate mass-flow-rate conditions and does not exhibit signs of extrapolative divergence. 

For sub-system 3, the model once again produces error distributions that are tightly clustered around zero and generally bounded within $-0.005$ m/s to $0.005$ m/s, with some data points extending beyond $0.005$ m/s. This consistent error confinement across both training and testing scenarios demonstrates the robustness of the model in capturing the corresponding velocity profiles, particularly under standard operating conditions, where flow features may be less complex. Overall, the results highlight the model’s strong predictive capability across the three sub-systems, with especially high accuracy observed for sub-systems 1 and 3 and acceptable, well-bounded performance for sub-system 2.

This behavior is consistent with the underlying model formulations used for each sub-system. Sub-systems 1 and 3 are governed by models derived directly from the fundamental conservation equations, which inherently encode the relevant physical dependencies and constraints. As a result, the learned mappings in these regions are more physically structured, leading to tightly concentrated error distributions and high predictive accuracy. In contrast, the model for sub-system 2 is based on a simplified algebraic relation rather than a first-principles formulation. This reduced physical expressiveness limits the model’s ability to capture more subtle variations in the flow field, which manifests as a broader spread in the prediction errors. Nevertheless, the errors for the intermediate (testing) mass-flow-rate conditions remain bounded by those of the training cases, indicating that the algebraic model still provides reliable predictions within its intended operating range.

Similar investigations are extended to extrapolative situations as well. {We evaluate the model’s capability to perform extrapolation beyond the range of inlet mass flow rates represented in the training data. Two extrapolative scenarios are considered: extrapolation toward lower mass flow rates and extrapolation toward higher mass flow rates.}

\textcolor{black}{For lower-end extrapolation, the model is trained exclusively on datasets corresponding to the intermediate and highest inlet mass flow rates (0.02970 kg/s and 0.03465 kg/s). The trained model is then tested on the lowest mass flow rate, 0.02475 kg/s, which lies outside the lower boundary of the training domain. This configuration assesses the model’s ability to reconstruct flow behavior when mass flow conditions are reduced below those observed during training.} \textcolor{black}{For higher-end extrapolation, the model is trained on the lowest and intermediate mass flow rates (0.02475 kg/s and 0.02970 kg/s) and subsequently tested on the highest mass flow rate, 0.03465 kg/s, which exceeds the upper bound of the training data. This scenario evaluates whether the model can generalize to more forceful extrusion conditions characterized by higher fluid momentum and elevated internal pressure.}

\begin{figure}[hbt!] % [h] places the figure approximately here
    \centering
    \includegraphics[width=0.95\textwidth]{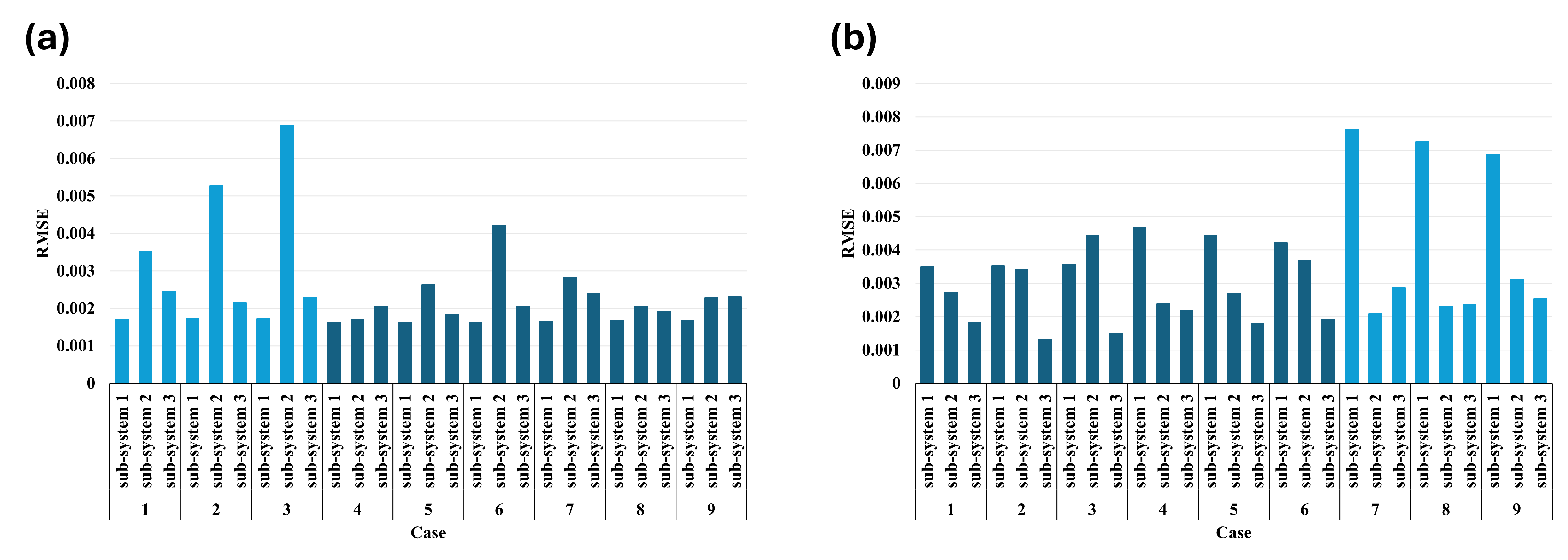} % Adjust the width as needed
    \caption{RMSE distribution of the dynamical model across all the datasets and sub-systems for the extrapolative mass flow rate training scenarios. (a) The dynamical model is trained with cases 4, 5, 6, 7, 8, and 9 and tested on cases 1, 2, and 3. (b) The dynamical model is trained with cases 1, 2, 3, 4, 5, and 6 and tested on cases 7, 8, and 9. Dark blue denotes the training cases, whereas light blue denotes the testing cases.}
    \label{fig:Ext_Low_High_RMSE}
\end{figure}

The RMSE distribution in \textcolor{black}{Figure \ref{fig:Ext_Low_High_RMSE}(a) illustrates the predictive performance of the model when trained on higher-end extrapolative mass flow rates. The results show improved prediction accuracy for all sub-systems compared to the interpolative training case, with sub-system 1 exhibiting particularly strong performance. This indicates that the model is capable of reliably extrapolating to slightly lower mass flow rates when trained on high-end data. In contrast, for sub-systems 2 and 3, the prediction errors are larger than both the training errors and those observed in the interpolative case. This increased error arises because these sub-systems are influenced by additional factors, such as surface velocity and coupled flow effects, that are not fully represented in the higher-end training data, limiting the model’s ability to extrapolate accurately in these regions.}  

\textcolor{black}{Similar trends are observed for the higher-end extrapolation scenario, as shown in Figure \ref{fig:Ext_Low_High_RMSE}(b). In general, the prediction errors for the high-end extrapolation are larger than those for the lower-end extrapolation. This is primarily because the model trained only on low mass flow rates, must extrapolate into a regime where the flow dynamics become nonlinear and coupled effects such as substrate-driven shear and momentum interactions are more pronounced. As a result, the model struggles to accurately capture the stronger velocities and pressure variations present at higher mass flow rates. In contrast, when the model is trained on high-flow cases, predicting lower-flow conditions is easier because the lower-flow dynamics can be regarded as a scaled-down version of the high-flow behavior. Consequently, predictive accuracy diminishes more significantly for high-end extrapolation than for low-end extrapolation.}  

\subsection{Model evaluation under different surface velocity conditions}

%\subsubsection{Interpolative}

%\textcolor{red}{training data: Case 1, 4, 7, 3, 6, 9}

\textcolor{black}{In this subsection, we first examine the model’s interpolative performance with respect to surface velocity. The model is trained using datasets representing the lowest and highest surface velocities in the simulation suite, at 50 mm/s and 70 mm/s, capturing the full range of nozzle–substrate relative motion. Training on these bounding cases exposes the model to both minimal and maximal shear and deformation effects imposed by the substrate motion.} \textcolor{black}{The trained model is subsequently tested on the intermediate surface velocity of 60 mm/s, which is excluded from the training data. This configuration allows us to evaluate how effectively the model predicts flow behavior for surface velocities that fall within the interior of the parameter space.}

\textcolor{black}{After interpolative scenario is evaluated, we investigate the model’s extrapolative performance with respect to surface velocity. In the first scenario, the model is trained on the intermediate and highest surface velocities, at 60 mm/s and 70 mm/s, which capture moderate to high relative motion between the nozzle and the build plate. These datasets provide the model with the dominant deformation and shear patterns that arise near the upper end of the operating range. The model is then tested on the lowest surface velocity of 50 mm/s, which lies outside the training region. This allows us to assess how well the model generalizes when confronted with deposition conditions characterized by reduced substrate motion and lower shear rates.}

\textcolor{black}{In the second scenario, we perform extrapolation toward higher velocity. The model is trained using datasets at 50 mm/s and 60 mm/s, which represent low to moderate surface velocities. These conditions primarily contain weaker shear interactions and smaller deformation gradients. The model is subsequently tested at 70 mm/s, a velocity that exceeds the training domain. This evaluation quantifies the model’s ability to generalize to more aggressive deposition conditions involving increased shear, higher strain rates, and larger material deformation.} 

%\subsubsection{Extrapolative}

%\textcolor{red}{training data: extrapolative 1: Case 2, 5, 8, 3, 6, 9}

%\textcolor{red}{training data: extrapolative 2: Case 1, 4, 7, 2, 5, 8}

\begin{figure}[hbt!] % [h] places the figure approximately here
    \centering
    \includegraphics[width=0.95\textwidth]{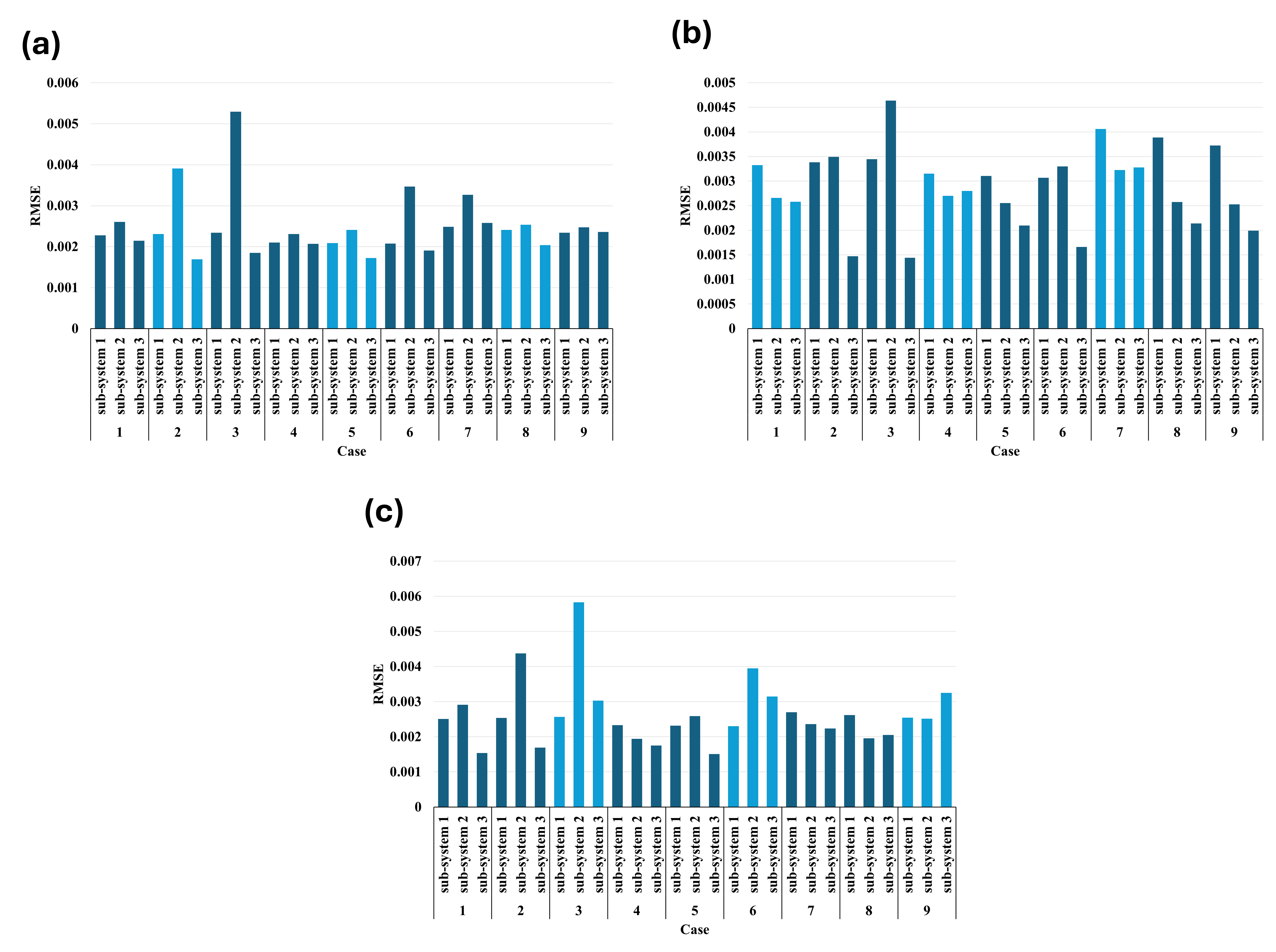} % Adjust the width as needed
    \caption{RMSE distribution of the dynamical model across all the datasets and sub-systems for different interpolative and extrapolative $U_s$ scenarios. (a) The interpolative scenario is trained  with cases 1, 3, 4, 6, 7, and 9 as the training cases and tested on cases 2, 5, and 8. (b) The low-end extrapolative scenario is trained  with cases 2, 3, 5, 6, 8, and 9 as the training cases and tested on cases 1, 4, and 7. (c) The high-end extrapolative scenario is trained  with cases 1, 2, 4, 5, 7, and 8 as the training cases and tested on cases 3, 6, and 9. Dark blue denotes the training cases, whereas light blue denotes the testing cases.}
    \label{fig:Int_Ext_Us_RMSE}
\end{figure}

The RMSE distributions for all scenarios are shown in Figure \ref{fig:Int_Ext_Us_RMSE}. The interpolative scenario, where the model predicts velocities within the range of training data, yields high accuracy, as expected, since the model operates within the domain it has already observed. These results are shown in shown in Figure \ref{fig:Int_Ext_Us_RMSE}(a). Minor deviations in the predictions are primarily due to the nonlinear interactions between mass flow and substrate motion, but overall the model captures the velocity profiles reliably.

\textcolor{black}{For the low-end extrapolation case, where the model is asked to predict flow at plate velocities below the training range, the predictions exhibit the highest accuracy among all scenarios. These results are shown in shown in Figure \ref{fig:Int_Ext_Us_RMSE}(b). This superior performance can be attributed to the fact that, in this system, lower plate velocities produce flow dynamics that are effectively a scaled-down version of those observed in the higher-velocity training cases. The model is able to capture the essential coupling between inlet mass flow and substrate-driven shear, and because the system behaves in a relatively linear or smoothly varying manner at lower velocities, extrapolation is easier.}

In contrast, the high-end extrapolation case (Figure \ref{fig:Int_Ext_Us_RMSE}(c)), where the model predicts velocities beyond the upper range of training plate velocities, exhibits the largest errors. This diminished accuracy arises from several factors. At higher plate velocities, the flow experiences stronger shear and more pronounced coupling with the substrate motion, introducing nonlinear effects that were not represented in the training data. The model, having no prior exposure to these extreme conditions, cannot fully account for the enhanced momentum transfer, entrainment, or deformation dynamics, leading to larger deviations from the ground truth. 

{Overall, these results indicate that the model generalizes best when extrapolating toward lower plate velocities and performs worst when required to extrapolate toward higher velocities, highlighting the asymmetry in extrapolative predictive capability due to nonlinear and coupled flow behaviors.}

\subsection{Model evaluation under randomized train–test partitioning}

\textcolor{black}{In addition to the systematic evaluations conducted across varying inlet mass flow rates and plate velocities, we further assess the model’s general predictive capability using a randomized train–test partitioning strategy. Unlike the previous two approaches with each designed to isolate the model’s sensitivity to specific operating parameters, this randomized sampling method provides a holistic measure of the model’s robustness when exposed to heterogeneous combinations of input conditions.}

\begin{figure}[hbt!] % [h] places the figure approximately here
    \centering
    \includegraphics[width=0.95\textwidth]{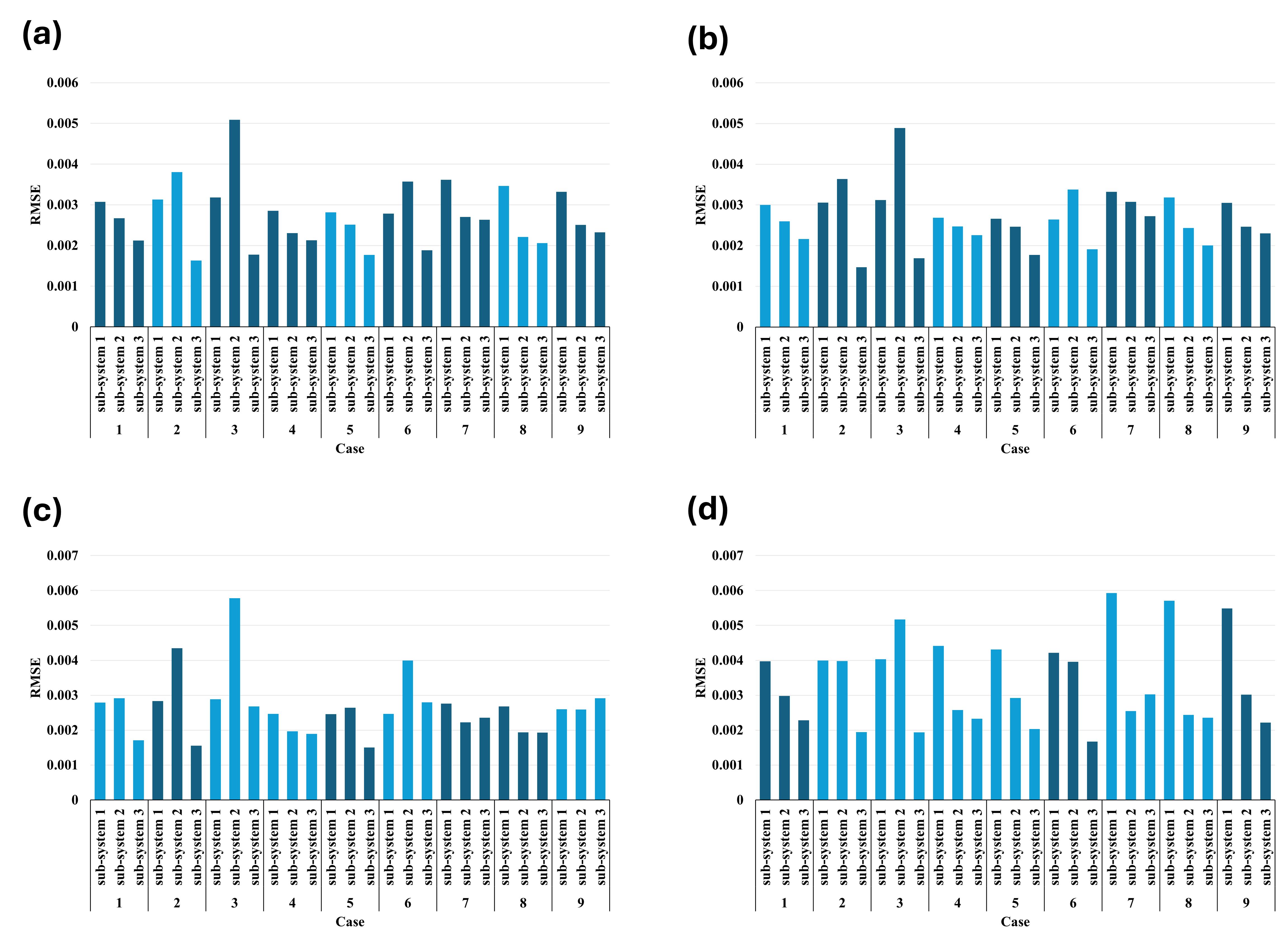} % Adjust the width as needed
    \caption{RMSE distribution of the dynamical model across all the datasets and sub-systems for random (a) 67-33, (b) 54-46, (c) 46-54, and (d) 33-67 training-testing splits. Dark blue denotes the training cases, whereas light blue denotes the testing cases.}
    \label{fig:Random_All_RMSE}
\end{figure}

\textcolor{black}{For this analysis, the complete dataset comprising all nine simulation cases is randomly divided into training and testing subsets using four different splits: 67–33, 56–44, 44–56, and 33–67 (Figure \ref{fig:Random_All_RMSE}). These splits correspond to 6 and 3, 5 and 4, 4 and 5, and 3 and 6 cases for training and testing, respectively. In the first split (67–33), cases 1, 3, 4, 6, 7, and 9 serve as the training set (Figure \ref{fig:Random_All_RMSE}(a)). In the second split (56–44), cases 2, 3, 5, 7, and 9 (Figure \ref{fig:Random_All_RMSE}(b)) serve as the training set. The third split (44–56) uses cases 2, 5, 7, and 8 for training (Figure \ref{fig:Random_All_RMSE}(c)), while the fourth split (33–67) uses cases 1, 6, and 9 for training (Figure \ref{fig:Random_All_RMSE}(d)). The model is trained on each training subset, and its predictive performance is then evaluated on the corresponding test cases. The resulting error metrics provide a quantitative assessment of the model’s generalization capability across heterogeneous operating conditions, offering a complementary overview that supplements the targeted extrapolative and interpolative evaluations discussed earlier.} 

\textcolor{black}{Across all four randomized train–test splits, the model demonstrates consistent predictive accuracy, with error levels comparable to those observed in the structured interpolative evaluations. As expected, splits with a larger proportion of training data (e.g., 67–33 and 56–44) yield the lowest prediction errors, reflecting the model’s ability to leverage broader exposure to heterogeneous operating conditions.}

{Even when the training sets combine low, moderate, and high mass flow rates with various plate velocities, the model maintains stable performance across all sub-systems, indicating that the reduced-order formulation effectively captures the dominant flow mechanisms irrespective of the specific training distribution. In contrast, the more restrictive splits with fewer training cases (44–56 and 33–67) show slightly elevated errors, yet no substantial degradation is observed, demonstrating that the model remains resilient even under limited data availability.} 

Across Figures \ref{fig:Random_All_RMSE}(b) to (d), in general, case 3 exhibits the highest error levels for sub-system 2, except for Figure \ref{fig:Random_All_RMSE}(d), where the training data are scarce. This behavior is physically expected, as case 3 corresponds to the lowest inlet mass-flow-rate condition combined with a relatively high plate velocity. Under these circumstances, the liquid film becomes more strongly influenced by shear-driven motion, leading to a thinner and more highly stretched film. Such flow regimes are more sensitive to small variations in operating conditions and therefore produce stronger nonlinear coupling between inertia, viscous effects, and wall shear.

When these enhanced shear-driven effects are modeled using the simplified algebraic formulation employed for sub-system 2, as opposed to the physics-based models used for sub-systems 1 and 3, the reduced physical expressiveness limits the model’s ability to fully capture the resulting flow behavior. This mismatch manifests as higher prediction errors, particularly in case 3 where film dragging is most pronounced. Consequently, the combination of a more challenging flow regime and the inherent limitations of the algebraic model explains the systematically increased errors observed for sub-system 2.

{Overall, these findings confirm that the proposed dynamical model generalizes reliably not only in controlled extrapolative and interpolative settings but also under randomized sampling, with improved accuracy naturally achieved when more training data are provided.}

\section{Conclusions}
\textcolor{black}{In this study, we developed a reduced-order, control-oriented dynamical model to describe the material flow behavior in extrusion-based 3D printing. The model comprises a set of differential–algebraic equations derived by partitioning the flow field into three regions: the nozzle interior, the nozzle–substrate gap, and the deposited layer on the moving build surface. Model parameters were identified using high-fidelity CFD simulation data, and the resulting model predictions were validated against an independent CFD testing dataset. The model achieves the highest accuracy within the nozzle region and effectively captures the dominant flow behavior in the gap and deposited-layer regions through a simplified linear representation that maintains acceptable error levels. Overall, the results demonstrate that the proposed dynamical model provides an accurate and computationally efficient foundation for future real-time control and optimization of extrusion-based 3D printing processes.}

\bibliographystyle{ieeetr}

\bibliography{ifacconf}             % bib file to produce the bibliography
                                                     % with bibtex (preferred)
                                                   
%\begin{thebibliography}{xx}  % you can also add the bibliography by hand

%\bibitem[Able(1956)]{Abl:56}
%B.C. Able.
%\newblock Nucleic acid content of microscope.
%\newblock \emph{Nature}, 135:\penalty0 7--9, 1956.

%\bibitem[Able et~al.(1954)Able, Tagg, and Rush]{AbTaRu:54}
%B.C. Able, R.A. Tagg, and M.~Rush.
%\newblock Enzyme-catalyzed cellular transanimations.
%\newblock In A.F. Round, editor, \emph{Advances in Enzymology}, volume~2, pages
%  125--247. Academic Press, New York, 3rd edition, 1954.

%\bibitem[Keohane(1958)]{Keo:58}
%\newblock \emph{Power and Interdependence: World Politics in Transitions}.
%\newblock Little, Brown \& Co., Boston, 1958.

%\bibitem[Powers(1985)]{Pow:85}
%\newblock Is there a way out?
%\newblock \emph{Harpers}, pages 35--47, June 1985.

%\bibitem[Soukhanov(1992)]{Heritage:92}
%\newblock \emph{{The American Heritage. Dictionary of the American Language}}.
%\newblock Houghton Mifflin Company, 1992.

%\end{thebibliography}

% \appendix
% \section{A summary of Latin grammar}    % Each appendix must have a short title.
% \section{Some Latin vocabulary}              % Sections and subsections are supported  
                                                                         % in the appendices.
\end{document}